\documentclass[prl,superscriptaddress,reprint,amsmath,amssymb,aps,longbibliography]{revtex4-1}
\usepackage{graphicx}
\usepackage{float}
\usepackage[
pdfstartview=FitH,
bookmarks=true,
breaklinks=true,
colorlinks=true,
linkcolor=blue,anchorcolor=blue,
citecolor=blue,filecolor=blue,
menucolor=blue,urlcolor=blue]{hyperref}

\usepackage{xcolor}
\usepackage{ulem}

\begin{document}

\title{Matter-wave Induced Transparency}

\author{Tongkang Wang}
\thanks{These authors contributed equally.}
\affiliation{Quantum Science Center of Guangdong-Hong Kong-Macao Greater Bay Area, Shenzhen 518045, China}
\affiliation{Department of Physics and State Key Laboratory of Low-Dimensional Quantum Physics, Tsinghua University, Beijing, 100084, China}
\author{Yuqi Liu}
\thanks{These authors contributed equally.}
\affiliation{Department of Physics and State Key Laboratory of Low-Dimensional Quantum Physics, Tsinghua University, Beijing, 100084, China}
\author{Wenlan Chen}
\email{cwlaser@ultracold.cn}
\affiliation{Department of Physics and State Key Laboratory of Low-Dimensional Quantum Physics, Tsinghua University, Beijing, 100084, China}
\author{Zhendong Zhang}
\email{zhendongzhang19950715@gmail.com}
\affiliation{Department of Physics and Hong Kong Institute of Quantum Science and Technology, The University of Hong Kong, Hong Kong, China}
\author{Jiazhong Hu}
\email{hujiazhong01@ultracold.cn}
\affiliation{Quantum Science Center of Guangdong-Hong Kong-Macao Greater Bay Area, Shenzhen 518045, China}

\begin{abstract}
Electromagnetically induced transparency suppresses optical absorption through destructive interference, playing a central role in light–matter interaction and quantum information science. We report matter-wave induced transparency, where atomic collisional interactions induce transmission through a lossy molecular potential for the incident atomic scattering waves. Using cesium Bose–Einstein condensates and modulation-induced Feshbach resonances, we realize a three-level atom–molecule coupled system with unprecedented flexibility. Under the dark state condition, a narrow and tunable transparency window appears within a broad dissipative collisional resonance. The transparency window linewidth is controlled by modulation-induced coupling. And scattering pathways are selectable via multifrequency Floquet modulation. These results establish an interference-based route for exploring programmable nonequilibrium and non-Hermitian physics, steering quantum chemistry and precision measurements.
\end{abstract}
\maketitle

\section*{Introduction}
Electromagnetically induced transparency (EIT) \cite{PhysRevLett.66.2593} is a paradigmatic example of quantum interference in light-matter interactions. In a driven three-level system, coherent coupling creates destructive interference that allows light to propagate through an otherwise opaque medium. This principle has enabled slow and stored light, quantum memories, enhanced optical nonlinearities, optical switching, and precision control of quantum states \cite{EITRev,Hau1999,science.1195596,EITApplication1,EITApplication2,EITApplication3}.

Matter waves obey the same principles of quantum superposition and interference as light, but they differ in one essential respect: they are intrinsically interacting. In ultracold gases, matter-wave interactions are often governed by collisions, where coherent scattering and inelastic loss are deeply intertwined. This raises a fundamental question: can transparency be induced directly by inherent collisional interactions, so that collisional loss is suppressed by interference between matter-wave scattering pathways rather than by interference between optical excitation pathways? 

Feshbach resonances \cite{Inouye_Andrews_Stenger_Miesner_Stamper-Kurn_Ketterle_1998,RevModPhys.78.1311,RevModPhys.82.1225,PhysRevA.70.032701,PhysRevA.76.042514,2012model} provide a natural setting for this question. A conventional magnetic Feshbach resonance can be viewed as an effective two-level matter-wave scattering system, in which an open-channel atomic pair couples to a closed-channel molecular state through collisional interactions. Near resonance, this coupling strongly modifies the scattering interaction, but the same molecular admixture can also open a dissipative pathway and enhance atom loss. The key step toward induced transparency is therefore to introduce additional scattering pathways that interfere destructively with this dissipative one, converting a single-pathway lossy resonance into a multichannel interference system.

\begin{figure*}[t]
    \centering
    \includegraphics[width=1\textwidth]{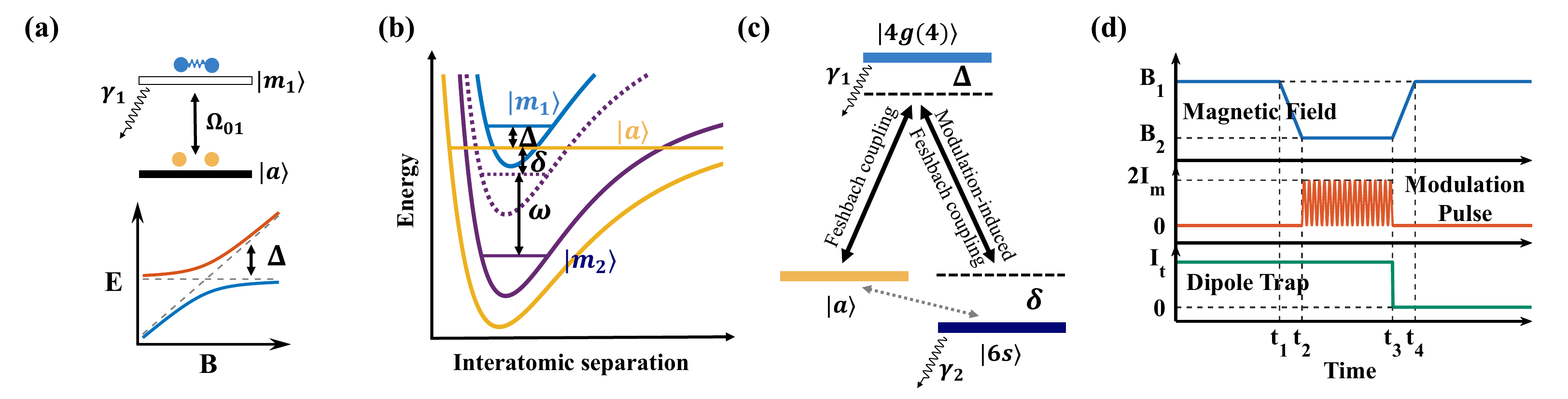}
    \caption{\textbf{From a two-level Feshbach resonance to a three-level matter-wave interference system.}
(a) A conventional magnetic Feshbach resonance can be viewed as an effective two-level system formed by an open-channel free-atom state \(|a\rangle\) and a closed-channel molecular state \(|m_1\rangle\). The magnetic field tunes the detuning \(\Delta\) between the two states, while the Feshbach coupling \(\hbar\Omega_{01}\) produces an avoided crossing. Near resonance, the admixture of the lossy molecular state (with decay rate \(\gamma_1\)) enhances inelastic collisional loss, analogous to absorption in a resonantly coupled two-level optical system.
(b) To construct a three-level system for interacting matter waves, a second molecular state \(|m_2\rangle\) in another closed-channel potential (solid purple curve) is introduced. Periodic energy modulation of the first molecular state \(|m_1\rangle\) (supported by the blue molecular potential curve) dresses \(|m_2\rangle\) with a Floquet energy quantum \(\hbar\omega\) (dashed purple curve), compensating the energy separation between \(|m_2\rangle\) and \(|m_1\rangle\). The residual detuning \(\delta\) controls the two-photon-like resonance condition required for matter-wave induced transparency.
(c) Effective \(\Lambda\)-type configuration realized with the free-atom state \(|a\rangle\) and two cesium Feshbach molecular states, \(|4g(4)\rangle\) (\(|m_1\rangle\)) and \(|6s\rangle\) (\(|m_2\rangle\)). The decay rates $\gamma_1$ and $\gamma_2$ describe the inelastic processes associated with these two molecular states (Supplementary Section II). The solid arrows denote the resonant couplings that form the effective three-level interference structure: the Feshbach coupling between \(|a\rangle\) and \(|4g(4)\rangle\), and the modulation-induced Feshbach coupling between \(|4g(4)\rangle\) and \(|6s\rangle\). 
These couplings lead to the interfering matter-wave scattering pathways responsible for the dark-state suppression of collisional loss.
The dashed arrow represents off-resonant background coupling that is present in the full Hamiltonian but is strongly suppressed in the effective resonant description (Supplementary Material).
(d) Experimental sequence. After preparing a cesium Bose--Einstein condensate, the magnetic field is ramped to the target value. A modulation pulse is then applied to engineer the desired three-level resonance condition. Finally, the modulation and dipole trap are switched off, and the remaining atom number is measured by absorption imaging after time of flight.}
    \label{fig1}
\end{figure*}

In contrast to previous optical or electromagnetic schemes that created destructive interference through externally driven transitions \cite{
PhysRevA.72.041801,PhysRevLett.95.063202,PhysRevLett.108.010401,PhysRevLett.116.075301}, we use matter-wave collisional coupling itself to create destructive interference, enabled by modulation-induced Feshbach resonances \cite{qrz6-gjp6,sciadv.adw3856,PhysRevLett.115.193002}. The additional pathways involved in this interference stem from a fundamental change in the structure of the scattering problem: the system is transformed from a two-level lossy Feshbach resonance into a three-level matter-wave interference structure that supports a dark-state condition. In this sense, matter-wave induced transparency (MWIT) realizes induced transparency of the collisional interaction itself, allowing dissipation to be suppressed while resonant interactions remain tunable. Such suppression requires a phase-coherent quantum description of scattering amplitudes and can not be captured by a classical rate-equation picture of collisional loss. Moreover, since MWIT harnesses pre-existing matter-wave interactions and avoids proximity to specific transitions, it allows for a lower-power drive field and weaker spontaneous emission.

In this work, we realize this transformation and observe MWIT in a cesium Bose--Einstein condensate. We couple a free-atom pair to two Feshbach molecular states and observe a narrow loss-suppressed window embedded within a strongly dissipative Feshbach resonance. The transparency position is tunable in the magnetic field--modulation frequency plane, the linewidth is controlled by the modulation-induced molecular coupling, and the scattering pathways can be selected by multifrequency Floquet modulation. This narrow transparency window and steep scattering-length dispersion may enable sensitive probes of magnetic fields and fundamental constants, as well as dispersion engineering of matter waves, such as controlling sound propagation in Bose--Einstein condensates. In addition, by separating resonant interactions from dissipation, MWIT provides a promising interference-based route to suppress reactive loss in ultracold molecules and engineer programmable dissipative dynamics. The same Floquet control of scattering pathways also enables resonance interference and bound states in the continuum, pointing to a general platform for controlling quantum interference in interacting matter waves.

\begin{figure*}[t]
    \centering
    \includegraphics[width=1\textwidth]{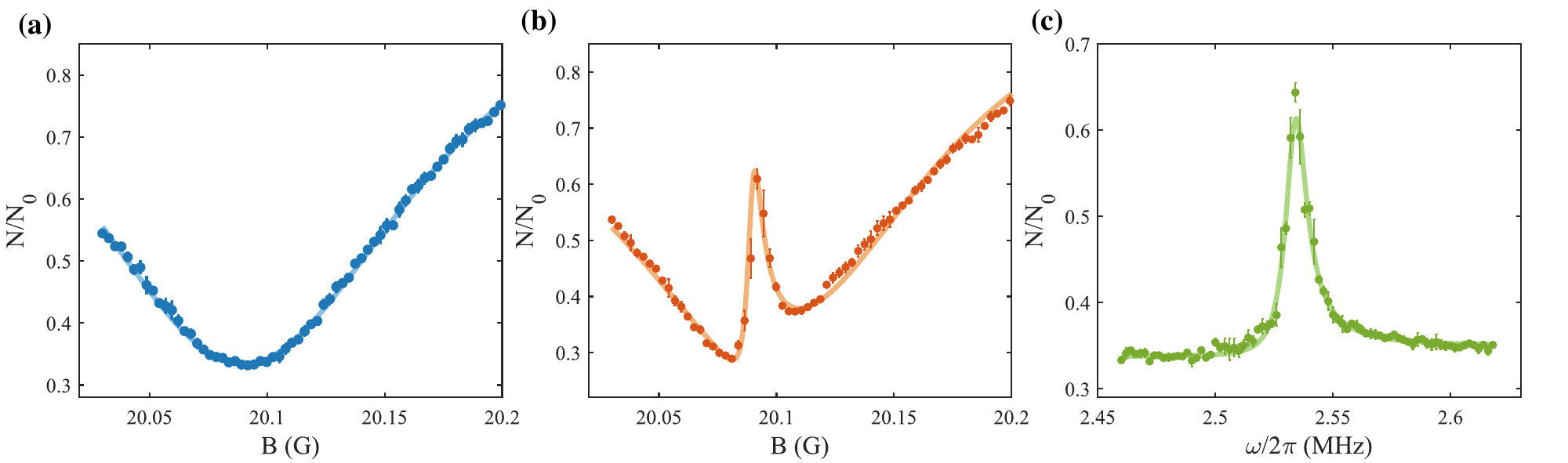}
    \caption{\textbf{Observation of matter-wave induced transparency in collisional loss.}
(a) Loss spectrum of a conventional two-level Feshbach resonance. The modulation frequency is fixed at \(\omega/2\pi = 1.5~\mathrm{MHz}\), for which the second molecular state is far from resonance (\(\delta/2\pi \simeq -1~\mathrm{MHz}\), where 1~MHz corresponds to an effective Zeeman shift at around 0.8~G), and the system is therefore treated as a two-level configuration. The remaining atom fraction \(N/N_0\) shows a broad loss feature associated with the first molecular state \(|m_1\rangle\). The center is shifted from the original \(|m_1\rangle\)-induced Feshbach resonance position of \(B_0=19.84~\mathrm{G}\) to \(B_0=20.09~\mathrm{G}\) due to the DC component light shift of the intensity-modulated pulse (Supplementary Material).
(b) Matter-wave transparency window in a strongly dissipative collisional resonance with a modulation frequency \(\omega/2\pi = 2.54~\mathrm{MHz}\). The second molecular state \(|m_2\rangle\) is brought close to the resonance condition \(\delta=0\) near $B_0$. A narrow loss-suppressed feature appears inside the broad Feshbach loss profile, indicating destructive interference among the matter-wave scattering pathways.
(c) Frequency-domain signature of the transparency resonance at a fixed magnetic field near the maximum-loss position of \(B_0=20.09~\mathrm{G}\). Scanning the modulation frequency reveals a pronounced increase of \(N/N_0\) around \(\omega/2\pi = 2.534~\mathrm{MHz}\), where the two-photon-like resonance condition \(\delta=0\) is satisfied. The solid curves are Fano-type fits to the measured loss spectra. All error bars denote the standard error of the mean.}
    \label{fig2}
\end{figure*}

\section*{Experimental platform and engineered three-level scheme}

Fig.~\ref{fig1}(a) illustrates a basic level structure underlying a conventional Feshbach resonance in a reduced
description of a two-level system: an open-channel free-atom state \(|a\rangle\) coherently couples to a closed-channel molecular state \(|m_1\rangle\) through inherent collisional interactions (e.g., isotropic electronic and relativistic spin-dependent interactions) \cite{RevModPhys.78.1311,RevModPhys.82.1225}. In this regard, a Feshbach resonance realizes an effective two-level coupled system, where the magnetic field \(B\) tunes the relative detuning \(\Delta\) between \(|a\rangle\) and \(|m_1\rangle\) (with \(|m_1\rangle\) referenced to \(|a\rangle\)). Plotted as an energy diagram versus \(B\), the resonance corresponds to a Landau–Zener–type avoided crossing between the two states, with a gap set by the Feshbach coupling. Near this avoided crossing, the admixture of the finite-lifetime molecular state \(|m_1\rangle\) into the free-atom state \(|a\rangle\) produces a resonant enhancement of interactions but also opens a dissipative pathway that leads to atom loss. The key ingredient required for induced transparency is an additional pathway that allows destructive interference within this lossy collisional dynamics.

We engineer such a pathway by coupling the molecular state \(|m_1\rangle\) to a longer-lived molecular state \(|m_2\rangle\) supported in another closed channel (Fig.~\ref{fig1}(b)), realized experimentally through a single-frequency periodic intensity-modulated pulse—the simplest case. 
This pulse modulates the energy level of $|m_1\rangle$ with a modulation frequency $\omega$ and compensates for the energy separation between \(|m_2\rangle\) and \(|m_1\rangle\) by a single drive quantum \(\hbar \omega\).
Simultaneously, the effective inter-channel coupling strengths are renormalized by their corresponding modulation factors (Supplementary Material) \cite{qrz6-gjp6,Supplementary}. The coupled molecular states \(|m_1\rangle\) and \(|m_2\rangle\), together with the free-atom state \(|a\rangle\), which couples to \(|m_1\rangle\), form an effective \(\Lambda\)-type three-level configuration (Fig.~\ref{fig1}(c)). The coupling between all states stems from the intrinsic matter-wave interaction, rather than from a direct optical transition. 
This is the key distinction from previous EIT or atom-molecule dark-state experiments based on electromagnetic or optical coupling.

Consequently, under the resonance condition \(\delta=0\) (Fig.~\ref{fig1}(b) and (c), analogous to a two-photon resonance), where \(\delta\) is the detuning of \(|m_2\rangle\) referenced to \(|a\rangle\), destructive interference generates a dark dressed state with reduced overlap on the lossy molecular component. This suppresses transfer from \(|a\rangle\) into decay channels and produces a matter-wave ''transparency window'' in the loss spectrum.

Fig.~\ref{fig1}(d) shows the experimental timing sequence. We start with a BEC of cesium-133 atoms in the $6S_{1/2}$ hyperfine ground state $|F=3,m_F=3\rangle$ with approximately $\mathrm{1 \times 10^5}$ atoms in a crossed optical dipole trap. A homogeneous magnetic field $B$ along the $z$ axis defines the quantization axis and tunes the energy difference between the atomic threshold and the Feshbach molecular states \cite{qrz6-gjp6}. After preparing the condensate, we ramp $B$ to a target value, apply a modulation pulse of duration $t$, and then measure the remaining atom number $N$ by absorption imaging after time of flight (Supplementary Material). Throughout this work, the primary observable is the remaining atom fraction $N/N_0$ (or equivalently the loss $1-N/N_0$), which directly reflects the inelastic collisions in our driven resonance. The experimental loss profiles as a function of magnetic field or modulation frequency are all fitted with a sum of Fano line shapes (Supplementary Material).

In our experiment, the molecular states \(|m_1\rangle\) and \(|m_2\rangle\) are chosen as the cesium Feshbach molecular states \(|4g(4)\rangle\) and \(|6s\rangle\), respectively \cite{PhysRevA.76.042514,2012model}. 
The dependence of the relevant detuning parameters on the magnetic field $B$ and modulation frequency $\omega$ is given by 
\begin{equation}
    \hbar\Delta = \Delta\mu_{1}(B-B_0)
    \label{Delta}
\end{equation}
and
\begin{equation}
    \hbar\delta = E_2(B_0)+\Delta\mu_{2}(B-B_0)+\hbar\omega,
    \label{delta}
\end{equation}
where $\Delta\mu_{1}$ is the magnetic moment difference between \(|4g(4)\rangle\) and \(|a\rangle\), $B_0$ is the Feshbach resonance position of \(|4g(4)\rangle\) under a modulation pulse with $\omega=0$, $E_2(B_0)$ is the energy of \(|6s\rangle\) referenced to \(|a\rangle\) at $B_0$ (also with $\omega=0$), and $\Delta\mu_{2}$ is the magnetic moment difference between \(|6s\rangle\) and \(|a\rangle\). Although the magnetic moment difference between \(|6s\rangle\) and \(|a\rangle\) generally varies with $B$, it can be approximated as constant within the experimental magnetic field range.

\begin{figure*}[t]
    \centering
    \includegraphics[width=1\textwidth]{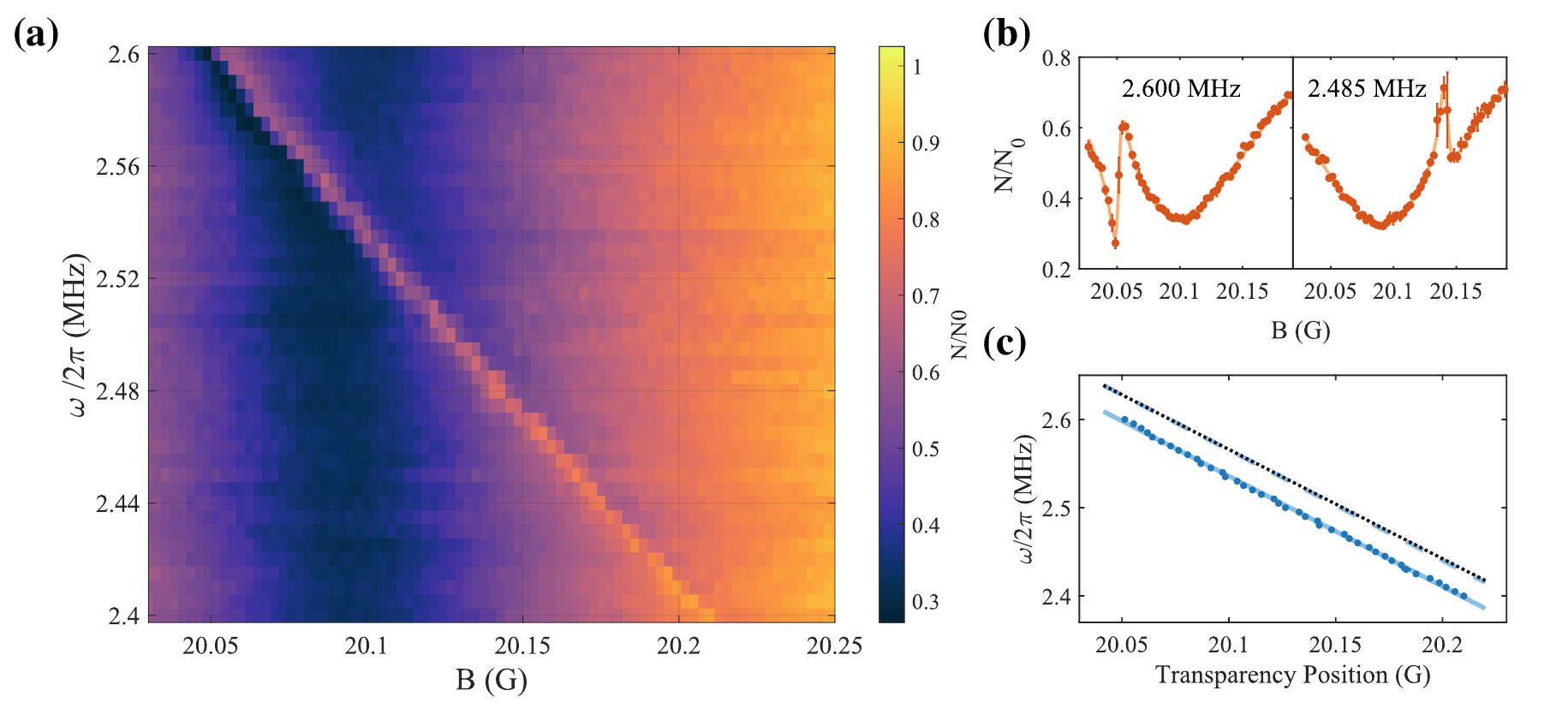}
    \caption{
\textbf{Dark-state condition of matter-wave induced transparency and resonance interference.}
(a) Two-dimensional map of the remaining atom fraction \(N/N_0\) as a function of magnetic field \(B\) and modulation frequency \(\omega\). The broad region of reduced atom number corresponds to the dissipative Feshbach resonance involving the lossy molecular state \(|m_1\rangle\). Within this lossy background, a narrow ridge of enhanced survival appears, marking the matter-wave transparency condition where the second molecular state \(|m_2\rangle\) participates in the three-level interference.
(b) Representative magnetic-field spectra taken at two different modulation frequencies, \(\omega/2\pi=2.600~\mathrm{MHz}\) and \(2.485~\mathrm{MHz}\). The loss-suppressed transparency feature appears at different magnetic fields for different modulation frequencies, demonstrating that the matter-wave transparency window can be continuously shifted across the dissipative Feshbach resonance. This tunability is analogous to conventional EIT, where the transparency condition is controlled by the frequency detuning of the coupling field.
The asymmetry between the two line shapes in (b) arises from a weak residual \(|a\rangle\leftrightarrow |m_2\rangle\) coupling and is a manifestation of resonance interference. This interference also produces a BIC in extended loss measurements across a broader range of modulation frequencies beyond that shown in (a) (Supplementary Sections IV and V~\cite{Supplementary}).
(c) The transparency position extracted from Fano-type fits to the magnetic-field spectra. The blue solid line represents the linear fit of the extracted Fano centers, while the blue dashed line corresponds to the fit including compensation for the light shift. The black dot-dashed line indicates the measured binding energy of the $|m_2\rangle$ state from modulation spectroscopy.
    }
    \label{fig3_3}
\end{figure*}

\section*{Observation of a transparency window in collisional loss}

We first measure the remaining atom fraction $N/N_0$ as a function of the magnetic field $B$ when the additional pathway is disabled (i.e., $\delta$ is far from resonance as $B$ is varied). This baseline loss spectrum is shown in Fig.~\ref{fig2}(a), exhibiting a conventional Lorentzian loss feature associated with a resonant two-level system.

In contrast, when the modulation frequency is set such that $\delta$ is near resonance as $\Delta$ approaches zero, the second scattering pathway is activated and a narrow transparency window emerges within the broader loss feature, as shown in Fig.~\ref{fig2}(b). The key observation is that the atom loss is suppressed at the center of an otherwise strongly dissipative Feshbach resonance, where loss would be maximal in the original two-level configuration. This suppression indicates destructive interference between matter-wave scattering pathways, which reduces the population of the lossy molecular component and forms a dark-state channel in collisions. The measured profile is well captured by one broad Fano line shape and one narrow Fano line shape, analogous to the EIT lineshape \cite{PhysRevA.81.053836,PhysRevLett.107.163604,Peng2014}, while here arising from intrinsic matter-wave coupling instead of direct optical transitions. A slight asymmetry in the narrow component is attributed to the weak coupling strength between $|a\rangle$ and $|m_2\rangle$ (Supplementary Material), which is denoted by the dashed arrow in the effective three-level configuration (Fig.~\ref{fig1}(c)). The emergence of a narrow loss-suppressed feature embedded within a broad dissipative resonance is the central signature of matter-wave induced transparency.

We also study the loss characteristics as a function of the modulation frequency at a fixed magnetic field near the resonance position $B_0$. The resulting experimental data in Fig.~\ref{fig2}(c) feature a loss suppression peak at the resonance frequency and are well fitted by a slightly asymmetric Fano profile (Supplementary Section IV~\cite{Supplementary}). 
This is analogous to an EIT measurement in which the probe detuning is fixed while the coupling-field detuning is scanned.

\begin{figure*}[t]
    \centering
    \includegraphics[width=1\textwidth]{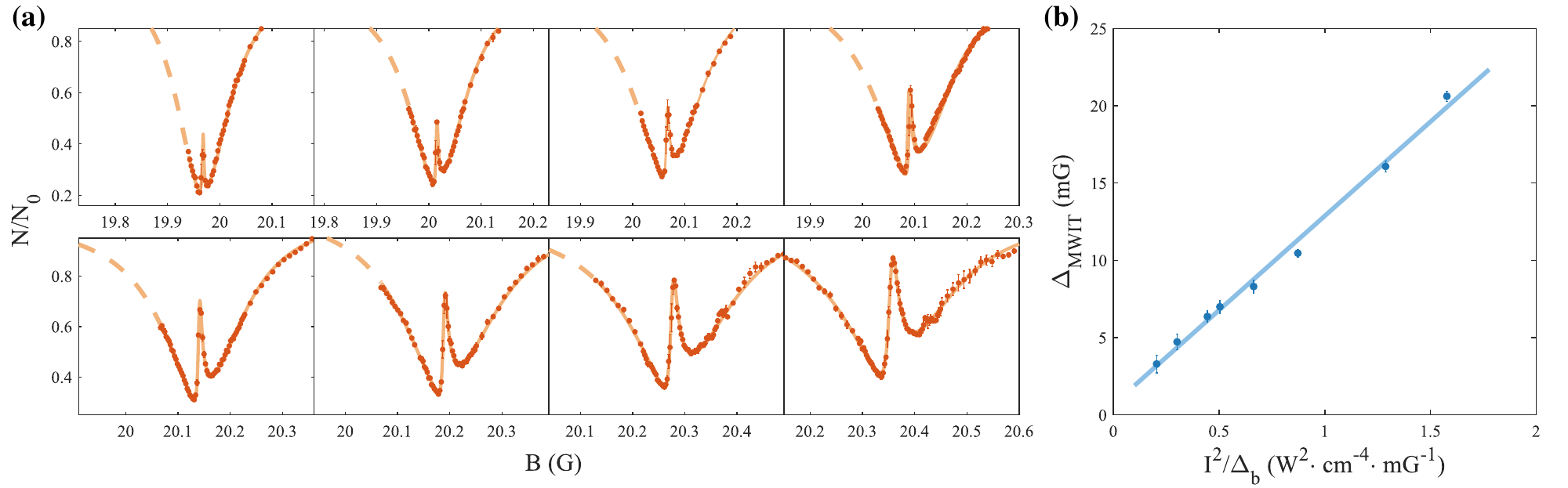}
    \caption{
\textbf{Coupling-controlled linewidth of matter-wave induced transparency.} (a) Magnetic-field loss spectra for increasing modulation intensity \(I\) (4.92, 6.65, 8.39, 9.66, 11.17, 13.82, 17.46, and 20.68~$\textrm{W/cm}^2$). As the modulation-induced collisional coupling is strengthened, the loss-suppressed transparency window becomes more pronounced and broadens within the dissipative Feshbach resonance. The dissipative loss features also broaden as inelastic processes strengthen (Supplementary Section V). The solid curves are Fano-type fits, and the dashed curves are extensions of these fits. (b) Narrow Fano branch linewidth \(\Delta_{\rm MWIT}\) of the transparency window extracted from the spectra in (a). The blue line is a linear fit to \(I^2/\Delta_{b}\).
    } 
    \label{fig4}
\end{figure*}

\section*{Dark state condition and Resonance interference}

To further connect the transparency to the relevant parameters, we perform a two-dimensional scan of the loss signal versus magnetic field $B$ and modulation frequency $\omega$. The results are shown in Fig.~\ref{fig3_3}(a) as a 2D colormap of the survival fraction $N/N_0$ in the $(B,\omega)$ plane.

Representative line cuts at two modulation frequencies are shown in Fig.~\ref{fig3_3}(b), where the modulation frequencies are chosen on opposite sides of the central frequency \(\omega_0\) satisfying \(\hbar\omega_0=E_2(B_0)\). 
The transparency feature shifts to different sides of the dissipative Feshbach resonance as the modulation frequency is changed. 
This behavior is analogous to conventional EIT, where the transparency window is displaced by tuning the coupling-field detuning; here, the relevant detuning is controlled by the modulation frequency that brings the second Feshbach molecular state into resonance. 
The two profiles, however, are not simply mirror-symmetric. 
Their asymmetric lineshapes reveal the presence of a weak residual \(|a\rangle\leftrightarrow |m_2\rangle\) coupling beyond the ideal \(\Lambda\)-type configuration, leading to interference between the dominant \(|m_1\rangle\)-induced lossy resonance and the weaker \(|m_2\rangle\)-associated resonance \cite{PhysRevA.32.3231}. 
Such interference between resonances can also give rise to bound states in the continuum (BICs) \cite{PhysRevA.32.3231,Hsu2016}. 
Motivated by this connection, we extend the loss measurement to a broader range of modulation frequencies. 
Near a specific modulation frequency, we observe the disappearance of one resonance branch, accompanied by a corresponding change in the loss profile, consistent with the calculated BIC condition for our coupled-resonance system (Supplementary Sections IV and V~\cite{Supplementary}).

In the $(B, \omega)$ plane, the transparency window appears as a narrow ridge of high survival probability along a well-defined curve, embedded within a broader region of strong loss. This curve is characterized by the transparency position $B(\omega)$, defined as the fitted magnetic field center of the narrow Fano branch at each modulation frequency (Supplementary Material). As shown in Fig.~\ref{fig3_3}(c), $B(\omega)$ is fitted using the expected relation

\begin{equation}
\hbar \omega=-E_2 (B_0) - \Delta\mu_2\,[B(\omega)-B_0] ,
\label{eq:resonance_condition}
\end{equation}
which corresponds to the dark state condition $\delta = 0$. The fitted value $\Delta\mu_2/h = 1.249(6)$ MHz/G is in good agreement with the measured value $1.238(8)~\mathrm{MHz/G}$ from modulation spectroscopy \cite{qrz6-gjp6}. After compensating the fitted line (blue solid line in Fig.~\ref{fig3_3}(c)) for the measured light shift from the DC component of the intensity-modulated pulse (Supplementary Section V~\cite{Supplementary}), the resulting line (blue dashed line in Fig.~\ref{fig3_3}(c)) matches the measured \(|m_2\rangle\) energy spectrum line (black dot-dashed line in Fig.~\ref{fig3_3}(c)). These observations demonstrate the dark state condition underlying matter-wave induced transparency.

\begin{figure*}[t]
    \centering
    \includegraphics[width=1.0\textwidth]{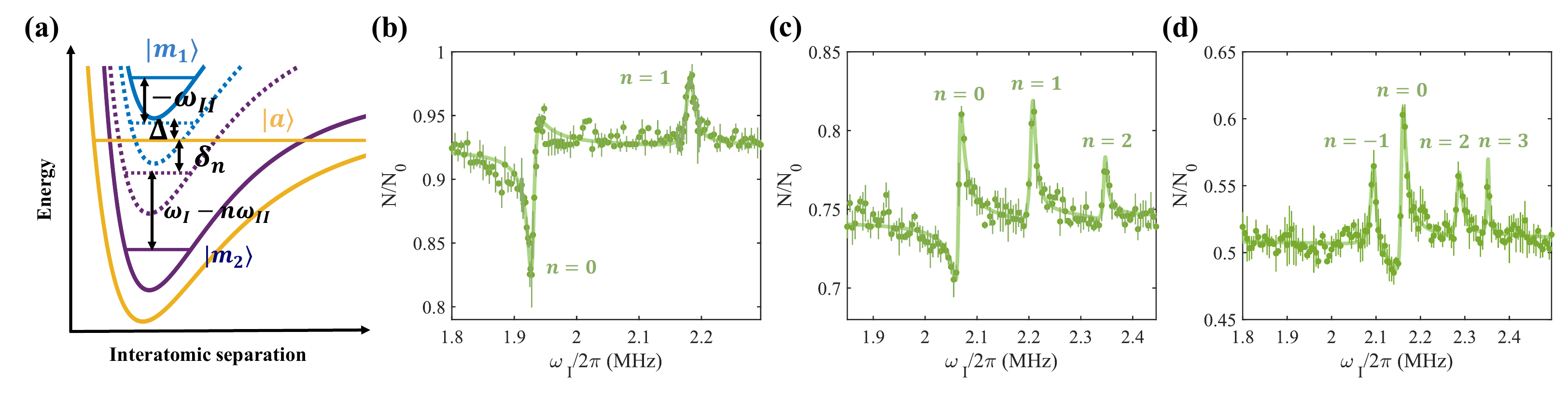}
    \caption{
\textbf{Multi-frequency engineering of matter-wave induced transparency.}
(a) In the dual-frequency modulation scheme, the dissipative background is created by setting the modulation tone \(\omega_{\mathrm{II}}\) to the first-order modulation-induced Feshbach resonance of the lossy molecular state \(|m_1\rangle\) ($\Delta=0$). Scanning the coupling tone \(\omega_{\mathrm{I}}\) brings the second molecular state \(|m_2\rangle\) into resonance with different Floquet sidebands labeled by $n$, each characterized by a two-photon-like detuning \(\delta_n\), resulting in different effective three-level coupling configurations.
(b) \(N/N_0\) as a function of the scanned modulation frequency \(\omega_{\mathrm{I}}\), with a second tone \(\omega_{\mathrm{II}}=2\pi\times 250\)~kHz fixed to create a dissipative \(|a\rangle\leftrightarrow |m_1\rangle\) background at \(B=20.57~\mathrm{G}\). A loss dip and a transparency peak appear. 
(c) At a different magnetic field \(B=20.44~\mathrm{G}\) with \(\omega_{\mathrm{II}}=2\pi\times 140\)~kHz, higher-order sidebands become visible. The features labeled by \(n=0,1,2\) demonstrate that the matter-wave transparency condition can be shifted and selected by the number of drive quanta exchanged with the fixed modulation tone.
(d) Sideband-selective suppression of collisional interference pathways. The second modulation frequency \(\omega_{\mathrm{II}}=2\pi\times 65\)~kHz with \(B=20.35~\mathrm{G}\). The absence of the \(n=1\) branch indicates that the corresponding effective coupling is suppressed by the Floquet dressing, while the remaining sidebands generate loss-suppressed transparency features. 
$B_0$ is shifted to 20.26 G for all three panels in this figure. For the second tone, the differential energy-modulation amplitude is 102 kHz between \(|m_1\rangle\) and \(|a\rangle\) and 155 kHz between \(|m_1\rangle\) and \(|m_2\rangle\).
}
    \label{Fig.5}
\end{figure*}

\section*{Intensity dependence and scaling of the transparency window}

We next investigate how the matter-wave transparency window depends on the modulation intensity \(I\), which controls the effective coupling between the two Feshbach molecular states.

Fig.~\ref{fig4}(a) shows magnetic-field loss spectra of \(N/N_0\) versus \(B\) for different modulation intensities \(I\) at a fixed modulation duration, with the modulation frequency chosen such that the transparency condition \(\delta \simeq 0\) is satisfied near the center of the dissipative Feshbach resonance. As \(I\) is increased, the modulation-induced coupling between \(|m_1\rangle\) and \(|m_2\rangle\) is strengthened. The loss-suppressed feature becomes more pronounced and broadens, while remaining embedded within the broad collisional loss profile associated with the lossy molecular state \(|m_1\rangle\). This behavior is the matter-wave analogue of the coupling-controlled transparency linewidth in EIT: stronger coupling opens a wider dark-state window inside an otherwise dissipative resonance.

In matter-wave induced transparency, this linewidth is controlled by modulation-induced Feshbach coupling (\(\Delta_{\mathrm{MWIT}} \propto |\Omega^{\mathrm{eff}}_{12}|^2 \)), for which the effective coupling strength between the relevant collisional states is approximately linear in modulation intensity in our experimental regime (i.e., \(\Omega^{\mathrm{eff}}_{12} \propto I\)).
To quantify this scaling, we extract the narrow Fano branch linewidth \(\Delta_{\mathrm{MWIT}}\) from fits of the measured loss spectra versus the magnetic field.
As shown in Fig.~\ref{fig4}(b), the extracted \(\Delta_{\mathrm{MWIT}}\) increases linearly with \(I^2/\Delta_{b}\) (analogous to power broadening in EIT; see Supplementary Section V~\cite{Supplementary}), where $\Delta_{b}$ is the Fano width of the measured broad dissipative Feshbach loss. 
The blue line is a linear fit to \(I^2/\Delta_{b}\), demonstrating a coupling-controlled linewidth,
\(\Delta_{\mathrm{MWIT}} \propto |\Omega^{\mathrm{eff}}_{12}|^2 \propto I^2\).

From the data in Fig.~\ref{fig4}(a), we also estimate the effective coupling strength \(|m_1\rangle\leftrightarrow |m_2\rangle\) by extracting the positions of the loss maxima and find agreement with the coupled-channel calculation (Supplementary Section V~\cite{Supplementary}).

\section*{Engineering with multi-frequency modulation}

To demonstrate that matter-wave induced transparency can be engineered beyond a single-frequency drive, we apply a dual-frequency modulation scheme. In this configuration (Fig.~\ref{Fig.5}(a)), a fixed modulation tone \(\omega_{\mathrm{II}}\) is tuned to the first-order modulation-induced Feshbach resonance of the lossy molecular state \(|m_1\rangle\), satisfying
\begin{equation}
\hbar\Delta \equiv \Delta\mu_1(B-B_0)-
\hbar\omega_{\mathrm{II}}=0.
\end{equation}
This tone resonantly couples the free-atom state \(|a\rangle\) to \(|m_1\rangle\) and thereby creates a controlled dissipative background. We then scan the coupling modulation tone \(\omega_{\mathrm{I}}\), which brings the second molecular state \(|m_2\rangle\) into resonance through different Floquet sidebands.

The corresponding two-photon-like detuning is rewritten as
\begin{equation}
\hbar\delta_n =
E_2(B_0)+\Delta\mu_2(B-B_0)
+\hbar(\omega_{\mathrm{I}}-n\omega_{\mathrm{II}}),
\end{equation}
where \(n\) labels the number of drive quanta exchanged with the fixed tone. When \(\delta_n=0\), the state \(|m_2\rangle\) participates in the collisional dynamics and an effective three-level interference structure can be formed. Because the effective couplings are dressed by Floquet Bessel factors, different sidebands can realize different coupling configurations: some enhance loss through an additional resonance, whereas others produce a loss-suppressed transparency peak.

Fig.~\ref{Fig.5}(b)-(d) show this sideband engineering of matter-wave induced transparency. In Fig.~\ref{Fig.5}(b), scanning \(\omega_{\mathrm{I}}\) reveals both a loss dip and a transparency peak. The dip corresponds to a sideband configuration in which the additional resonance does not form an efficient dark-state pathway, leading to enhanced dissipation. In contrast, the peak corresponds to a sideband where the \(|m_1\rangle\leftrightarrow |m_2\rangle\) coupling is active while the direct \(|a\rangle\leftrightarrow |m_2\rangle\) coupling is suppressed, producing matter-wave induced transparency. As the magnetic field is tuned closer to \(B_0\), the required \(\omega_{\mathrm{II}}\) decreases and higher-order Floquet sidebands become visible, as shown in Fig.~\ref{Fig.5}(c). The observed features correspond to different values of \(n\), demonstrating that the transparency condition can be selected by the modulation sideband.

The flexibility of this approach is further illustrated in Fig.~\ref{Fig.5}(d), where one expected sideband branch is strongly suppressed. This missing branch ($n=1$) arises from the simultaneous suppression of the relevant effective couplings by the modulation-induced Bessel factors (Supplementary Material). The remaining features correspond to other allowed Floquet sidebands and exhibit loss-suppressed transparency. These results show that multi-frequency modulation provides a programmable way to select, enhance, or suppress atom--molecule scattering pathways, thereby engineering the multilevel interference responsible for matter-wave induced transparency.

\section*{Discussion and outlook}

We have demonstrated matter-wave induced transparency by realizing a controllable three-level quantum-interference system via modulation-induced Feshbach resonances. In contrast to conventional EIT, where electromagnetic or optical coupling creates destructive interference that suppresses optical absorption, the interference that suppresses collisional loss in MWIT is induced by intrinsic atomic collisional coupling and occurs among interacting matter-wave scattering pathways.
The observation of a transparency window embedded in a dissipative Feshbach resonance (Fig.~\ref{fig2}), its tunable transparency position in the \((B,\omega)\) plane (Fig.~\ref{fig3_3}), the coupling-controlled linewidth scaling (Fig.~\ref{fig4}), and the sideband-selective engineering enabled by multi-frequency modulation (Fig.~\ref{Fig.5}) together establish induced transparency as a robust interference phenomenon in collisional quantum matter.

This result points to a general strategy for separating resonant interactions from dissipation. In ultracold atoms and molecules, access to strongly interacting regimes is often limited by inelastic loss, chemical reactions, or long-lived collision complexes. Matter-wave induced transparency provides an interference-based route to suppress such loss while retaining coherent control over resonant scattering. This capability is particularly relevant for ultracold molecular gases, where quantum interference has already emerged as a powerful tool for controlling reactive collisions and few-body pathways\cite{Reaction1,Reaction2,Efimov3}. 
Extending the present mechanism to molecular systems may therefore enable longer-lived strongly interacting gases and controlled quantum chemistry. 

Beyond loss suppression, the narrow transparency window provides a promising tool for precision measurements of the magnetic field and may be sensitive to variations in fundamental constants~\cite{PhysRevLett.96.230801}. Given the analogy between phonon in BEC and photon in EIT, the steep scattering-length dispersion of MWIT (Supplementary Section IV) could be used to engineer nontrivial phonon velocity fields, opening new possibilities for cosmological simulations, such as simulating gravitational fields in BEC~\cite{Viermann2022}. It may also be used to reduce the group velocity of colliding Bose–Einstein condensates\cite{Group}, with potential applications in matter-wave quantum memories.

More broadly, the modulation approach developed here offers a flexible Floquet toolbox for selecting, enhancing, or suppressing specific scattering pathways, enabling the observation of new quantum phenomena, including not only matter-wave induced transparency but also resonance interference and a novel class of bound states in the continuum (BICs) in our experiment. Although all of these observations are based on a three-level structure, the same approach can be used to manipulate interference in other configurations. For example, we also find that a modulated two-level matter-wave system enables the control of Landau–Zener–St\"uckelberg–Majorana interference \cite{LZSMReview}, which may lead to a new type of matter-wave interferometer and will be explored in our future work. In general, this method could be applied to programmable non-Hermitian quantum physics and nonequilibrium many-body dynamics in driven quantum gases and other related platforms.

\section*{ACKNOWLEDGMENTS}
We would like to thank Prof. Cheng Chin for stimulating discussions. This work is supported by the Quantum Science and Technology-National Science and Technology Major Project, the Guangdong Provincial Quantum Science Strategic Initiative, the National Natural Science Foundation of China, and the National Key Research and Development Program of China.

\bibliography{reference.bib}

\end{document}




\title{Supplementary materials for Matter-wave Induced Transparency}

\author{Tongkang Wang}
\thanks{These authors contributed equally.}
\affiliation{Quantum Science Center of Guangdong-Hong Kong-Macao Greater Bay Area, Shenzhen 518045, China}
\affiliation{Department of Physics and State Key Laboratory of Low-Dimensional Quantum Physics, Tsinghua University, Beijing, 100084, China}
\author{Yuqi Liu}
\thanks{These authors contributed equally.}
\affiliation{Department of Physics and State Key Laboratory of Low-Dimensional Quantum Physics, Tsinghua University, Beijing, 100084, China}
\author{Wenlan Chen}
\email{cwlaser@ultracold.cn}
\affiliation{Department of Physics and State Key Laboratory of Low-Dimensional Quantum Physics, Tsinghua University, Beijing, 100084, China}
\author{Zhendong Zhang}
\email{zhendongzhang19950715@gmail.com}
\affiliation{Department of Physics and Hong Kong Institute of Quantum Science and Technology, The University of Hong Kong, Hong Kong, China}
\author{Jiazhong Hu}
\email{hujiazhong01@ultracold.cn}
\affiliation{Quantum Science Center of Guangdong-Hong Kong-Macao Greater Bay Area, Shenzhen 518045, China}
\maketitle


\begingroup
\renewcommand{\baselinestretch}{1.2}\normalsize
\tableofcontents
\endgroup

\newpage
\section{EXPERIMENTAL METHODS}
\subsection{Experimental sequence} The experiment starts with an almost pure Bose--Einstein condensate of \(N\simeq 1\times10^5\) cesium-133 atoms in the \(6S_{1/2}\) hyperfine ground state \(|F=3,m_F=3\rangle\). 
The atoms are confined in a horizontally crossed optical dipole trap formed by 1064-nm laser beams. The initial magnetic field is set to \(20.43~\mathrm{G}\), where the scattering length is \(a=200a_0\). The trap vibrational frequencies along the three directions are \((\omega_x,\omega_y,\omega_z)=2\pi\times(22,23,56)~\mathrm{Hz}\).

We ramp the current of magnetic field coil to the target value in 0.3~ms and then wait for 3~ms so that the magnetic-field contribution from eddy currents becomes negligible. A modulation pulse is then applied. The modulation is synchronized to the 50-Hz line signal to reduce the effect of magnetic-field ripple. We define the atom number at the beginning of the modulation pulse as 
\(N_0\). For most magnetic-field values, \(N_0\) is equal to the initially prepared atom number. An exception occurs near the narrow \(g\)-wave magnetic Feshbach resonance at \(19.84~\mathrm{G}\), where atom loss can take place before the modulation pulse is applied. Before taking data at a given parameter set, we first perform a control sequence without a modulation pulse to identify the magnetic-field range that produces such background loss, and then we avoid this range in the measurements.

After a 2~ms modulation pulse, we switch off the modulation light and the dipole trap, and ramp the magnetic field back to 20.43~G for expansion and imaging. The remaining atom number is measured by absorption imaging after a 17.5~ms time of flight.

\subsection{Magnetic field calibration} We calibrate the magnetic field using the microwave transition from 
\(|F=3,m_F=3\rangle\) to \(|F=4,m_F=4\rangle\) in cesium, where \(F\) and 
\(m_F\) denote the total angular momentum and the magnetic quantum number, 
respectively. With synchronization to the 50-Hz line signal, the magnetic field 
is stable to within 4~mG.

\subsection{Modulation protocol} 
The underlying mechanism of our modulation scheme can be captured by a simplified three-level model. 
The corresponding coupled-channel treatment is provided in the Supplementary Section~\ref{three-channel}. 
We consider three states, denoted by \(|a\rangle\), \(|m_1\rangle\), and \(|m_2\rangle\). 
Here \(|a\rangle\) is the free-atom state, while \(|m_1\rangle\) and \(|m_2\rangle\) are Feshbach molecular states.

We start from the time-dependent Hamiltonian
\begin{equation}
H_0(t)=\hbar
\begin{pmatrix}
E_0(t)/\hbar & \Omega_{01}/2 & \Omega_{02}/2 \\
\Omega_{01}^{*}/2 & E_1(t)/\hbar & \Omega_{12}/2 \\
\Omega_{02}^{*}/2 & \Omega_{12}^{*}/2 & E_2(t)/\hbar
\end{pmatrix},
\end{equation}
where
\begin{eqnarray}
\frac{E_i(t)}{\hbar}
&=&
\frac{E_i^0}{\hbar}
+\alpha_i\cos(\omega_{\rm I}t+\phi_{\rm I})
\nonumber \\ 
& &+\beta_i\cos(\omega_{\rm II}t+\phi_{\rm II}).
\end{eqnarray}
Here \(i=0,1,2\) corresponds to \(|a\rangle\), \(|m_1\rangle\), and \(|m_2\rangle\), respectively. 
The parameters \(E_i^0\) are the unmodulated energies. 
The parameters \(\alpha_i\) and \(\beta_i\) are the modulation amplitudes induced by the two modulation tones \(\omega_{\rm I}\) and \(\omega_{\rm II}\), respectively, and \(\phi_{\rm I}\) and \(\phi_{\rm II}\) are their phases. 
The coupling \(\Omega_{ij}\) denotes the intrinsic coupling between states \(i\) and \(j\), arising from interatomic interactions such as exchange and van der Waals interactions, as well as magnetic dipole interaction and second-order spin-orbit coupling, which are important for cesium atoms~\cite{Method1,2012model}.

We then apply the unitary transformation
\begin{equation}
U(t)=
\begin{pmatrix}
U_{00}(t) & 0 & 0 \\
0 & U_{11}(t) & 0 \\
0 & 0 & U_{22}(t)
\end{pmatrix},
\end{equation}
with
\begin{eqnarray}
U_{00}(t)
&=&
\exp\left\{
-i\left[
\frac{E_0^0}{\hbar}t
+\frac{\alpha_0}{\omega_{\rm I}}\sin(\omega_{\rm I}t+\phi_{\rm I})\right.\right.
 \nonumber \\
 & &\left.\left.+\frac{\beta_0}{\omega_{\rm II}}\sin(\omega_{\rm II}t+\phi_{\rm II})
\right]\right\},
\nonumber\\
U_{11}(t)
&=&
\exp\left\{
-i\left[
\frac{E_0^0}{\hbar}t
+l\omega_{\rm II}t
+\frac{\alpha_1}{\omega_{\rm I}}\sin(\omega_{\rm I}t+\phi_{\rm I})
\right.\right.\nonumber \\
& &\left.\left.+\frac{\beta_1}{\omega_{\rm II}}\sin(\omega_{\rm II}t+\phi_{\rm II})
\right]\right\},
\nonumber\\
U_{22}(t)
&=&
\exp\left\{
-i\left[
\frac{E_0^0}{\hbar}t
+m\omega_{\rm I}t
+n\omega_{\rm II}t \right.\right.\nonumber \\
& &\left.\left.
+\frac{\alpha_2}{\omega_{\rm I}}\sin(\omega_{\rm I}t+\phi_{\rm I})
+\frac{\beta_2}{\omega_{\rm II}}\sin(\omega_{\rm II}t+\phi_{\rm II})
\right]\right\}.
\end{eqnarray}
Here \(l\), \(m\), and \(n\) are integers labeling the Floquet sidebands. 
The transformed Hamiltonian is
\begin{equation}
H(t)=U^\dagger(t)H_0(t)U(t)-i\hbar U^\dagger(t)\frac{\partial U(t)}{\partial t}.
\end{equation}
It can be written as
\begin{equation}
H(t)=\hbar
\begin{pmatrix}
H_{00} & H_{01} & H_{02} \\
H_{01}^{*} & H_{11} & H_{12} \\
H_{02}^{*} & H_{12}^{*} & H_{22}
\end{pmatrix},
\end{equation}
where
\begin{eqnarray}
H_{00}
&=&0,
\nonumber\\
H_{11}
&=&\frac{E_1^0-E_0^0}{\hbar}-l\omega_{\rm II},
\nonumber\\
H_{22}
&=&\frac{E_2^0-E_0^0}{\hbar}-m\omega_{\rm I}-n\omega_{\rm II},
\end{eqnarray}
and
\begin{eqnarray}
H_{01}
&=&
\frac{\Omega_{01}}{2}
\exp\left\{
-i\left[
l\omega_{\rm II}t
+\frac{\alpha_1-\alpha_0}{\omega_{\rm I}}\sin(\omega_{\rm I}t+\phi_{\rm I})
\right.\right.\nonumber \\
& &\left.\left.+\frac{\beta_1-\beta_0}{\omega_{\rm II}}\sin(\omega_{\rm II}t+\phi_{\rm II})
\right]\right\},
\nonumber\\
H_{02}
&=&
\frac{\Omega_{02}}{2}
\exp\left\{
-i\left[
m\omega_{\rm I}t+n\omega_{\rm II}t
+\frac{\alpha_2-\alpha_0}{\omega_{\rm I}}\sin(\omega_{\rm I}t+\phi_{\rm I})
\right.\right.\nonumber \\
& &\left.\left.+\frac{\beta_2-\beta_0}{\omega_{\rm II}}\sin(\omega_{\rm II}t+\phi_{\rm II})
\right]\right\},
\nonumber\\
H_{12}
&=&
\frac{\Omega_{12}}{2}
\exp\left\{
-i\left[
m\omega_{\rm I}t+(n-l)\omega_{\rm II}t
+\frac{\alpha_2-\alpha_1}{\omega_{\rm I}}
\right.\right.\nonumber \\
& &\left.\left.
\times\sin(\omega_{\rm I}t+\phi_{\rm I})
+\frac{\beta_2-\beta_1}{\omega_{\rm II}}\sin(\omega_{\rm II}t+\phi_{\rm II})
\right]\right\}.
\end{eqnarray}

Using the Jacobi--Anger expansion
\begin{eqnarray}
e^{ix\sin\theta}
&=&
\sum_{k=-\infty}^{\infty}J_k(x)e^{ik\theta},
\end{eqnarray}
where \(J_k(x)\) is the \(k\)-th order Bessel function of the first kind, and keeping only the near-resonant terms under the rotating-wave approximation, the Hamiltonian is simplified to
\begin{equation}
H_{\rm eff}
=
\hbar
\begin{pmatrix}
0 & \Omega_{01}^{\rm eff}/2 & \Omega_{02}^{\rm eff}/2 \\
(\Omega_{01}^{\rm eff})^{*}/2 & \Delta & \Omega_{12}^{\rm eff}/2 \\
(\Omega_{02}^{\rm eff})^{*}/2 & (\Omega_{12}^{\rm eff})^{*}/2 & \delta
\end{pmatrix}.
\end{equation}
Here
\begin{eqnarray}
\Delta
&=&
\frac{E_1^0-E_0^0}{\hbar}-l\omega_{\rm II},
\nonumber\\
\delta
&=&
\frac{E_2^0-E_0^0}{\hbar}-m\omega_{\rm I}-n\omega_{\rm II}.
\end{eqnarray}
The effective couplings are
\begin{eqnarray}
\Omega_{01}^{\rm eff}
&=&
\Omega_{01}
J_0\left(\frac{\alpha_0-\alpha_1}{\omega_{\rm I}}\right)
J_l\left(\frac{\beta_0-\beta_1}{\omega_{\rm II}}\right)
e^{il\phi_{\rm II}},
\nonumber\\
\Omega_{02}^{\rm eff}
&=&
\Omega_{02}
J_m\left(\frac{\alpha_0-\alpha_2}{\omega_{\rm I}}\right)
J_n\left(\frac{\beta_0-\beta_2}{\omega_{\rm II}}\right)
e^{i(m\phi_{\rm I}+n\phi_{\rm II})},
\nonumber\\
\Omega_{12}^{\rm eff}
&=&
\Omega_{12}
J_m\left(\frac{\alpha_1-\alpha_2}{\omega_{\rm I}}\right)
\times \nonumber \\
& &J_{n-l}\left(\frac{\beta_1-\beta_2}{\omega_{\rm II}}\right)
e^{i[m\phi_{\rm I}+(n-l)\phi_{\rm II}]}.
\end{eqnarray}

When \(\beta_0-\beta_1=0\), \(\beta_0-\beta_2=0\), 
\(\alpha_0-\alpha_2=0\), \(l=0\), \(n=0\), and \(m=-1\), 
we obtain an effective Hamiltonian
\begin{equation}
H_{\rm eff}
=
\hbar
\begin{pmatrix}
0 & \Omega_{01}^{\rm eff}/2 & 0 \\
(\Omega_{01}^{\rm eff})^{*}/2 & \Delta & \Omega_{12}^{\rm eff}/2 \\
0 & (\Omega_{12}^{\rm eff})^{*}/2 & \delta
\end{pmatrix}.
\end{equation}
This corresponds to the ideal single-frequency modulation case for matter-wave induced transparency. 
The effective direct coupling between \(|a\rangle\) and \(|m_2\rangle\) is suppressed because 
\[
J_{-1}\left(\frac{\alpha_0-\alpha_2}{\omega_{\rm I}}\right)=0
\]
under the above condition \(\alpha_0-\alpha_2=0\). 
The remaining two couplings, \(\Omega_{01}^{\rm eff}\) and \(\Omega_{12}^{\rm eff}\), form an effective \(\Lambda\)-type configuration involving the free-atom state \(|a\rangle\) and the two Feshbach molecular states \(|m_1\rangle\) and \(|m_2\rangle\). 
This Hamiltonian has the same structure as the standard three-level Hamiltonian used to describe EIT, but here the couplings originate from atom--molecule collisional coupling rather than optical transitions.

If \(\alpha_0-\alpha_2\) is nonzero but small, a weak direct coupling 
\(\Omega_{02}^{\rm eff}\) remains in the effective Hamiltonian. 
This residual coupling accounts for the dashed line in Fig.~1(c) in the main text and contributes to the asymmetric Fano lineshapes observed in the experiment.

For the dual-frequency modulation measurements shown in Fig.~5 in the main text, we choose \(l=1\) and \(m=-1\). 
In this case, the fixed modulation tone \(\omega_{\rm II}\) resonantly couples 
\(|a\rangle\) to \(|m_1\rangle\), while the scanned tone \(\omega_{\rm I}\), together with different sideband indices \(n\), brings \(|m_2\rangle\) into resonance. 
The same effective Hamiltonian therefore describes the multi-frequency engineering of the three-level matter-wave interference system, with the effective couplings controlled by the corresponding Bessel-function factors. For the missing $n=1$ branch, $\Omega_{12}^{\rm eff}\simeq0$ since $J_{0}(\frac{155~\mathrm{kHz}}{65~\mathrm{kHz}})\simeq0$, and $\Omega_{02}^{\rm eff}$ is further suppressed by an additional Bessel function $J_1$ compared to the single-tone case.

A detailed discussion of the effects of various effective three-level configurations under coupled-channel treatment is provided in the Supplementary Section~\ref{section2}.

In the experiment, a far-detuned laser shifts the energies of the free-atom state and Feshbach molecular states through the AC Stark effect. Intensity modulation of this laser provides the time-dependent energy modulation described by \(\alpha_i\) and \(\beta_i\) in the theoretical model. More experimental details are provided in Supplementary Section~\ref{section4}.

\subsection{Loss data fitting} 
All atom-loss data as a function of magnetic field \(B\) or modulation frequency 
are fitted with a sum of Fano line shapes \cite{Fano,Method4},
\begin{equation}
\frac{N}{N_0}
=
c_0+\sum_i c_i\frac{(q_i+\epsilon_i)^2}{1+\epsilon_i^2}.
\end{equation}
When fitting magnetic-field spectra, \(\epsilon_i=2(B-B_i)/\Delta_i\);
when fitting frequency spectra, \(\epsilon_i=2(\omega-\omega_i)/\Gamma_i\).
Here \(q_i\) is the Fano parameter, \(c_i\) is the amplitude, \(B_i\) or \(\omega_i\) is the Fano center, \(\Delta_i\) or \(\Gamma_i\) is the linewidth, and \(c_0\) is a constant offset.

For the magnetic-field spectra in Fig.~2,~3 and~4 in the main text, the measured profiles are fitted with one broad Fano line shape and one narrow Fano line shape (i.e. $i\in\{1,2 \}$). The transparency position is determined from the fitted narrow branch Fano center. The linewidth of the transparency window is extracted as the fitted narrow Fano branch linewidth.

The loss profiles in the frequency spectra of Figs.~2 and~5 in the main text are fitted with one or multiple Fano profiles, depending on the number of observed peaks or dips.

\section{THE SCATTERING PROPERTIES OF MATTER-WAVE INDUCED TRANSPARENCY}
\label{three-channel}

{\subsection{Effective Hamiltonian Under Modulation}} 
To explain the phenomenon observed in matter-wave induced transparency, we employ a simplified three-channel model similar to that used in modulation-induced Feshbach resonances~\cite{Ref1}. 
The three channels are an open atomic scattering channel \(\ket{\rm op}\), a closed molecular channel \(\ket{\rm cl_1}\), and a second closed molecular channel \(\ket{\rm cl_2}\). The open-channel threshold is set to zero energy, and the atomic scattering wave is incident from the open channel. \(\ket{\rm cl_1}\) and \(\ket{\rm cl_2}\) support molecular states \(\ket{m_1}\) and \(\ket{m_2}\) with energy $E_1^{0}$ and $E_2^{0}$, respectively.
The open channel is coupled to \(\ket{\rm cl_1}\) and \(\ket{\rm cl_2}\) through the intrinsic channel couplings $W_1$ and $W_2$, respectively, while the two closed channels are coupled by $V$. When dual-frequency modulation is applied to the energy of these channels, the resulting time-dependent Hamiltonian $H_0(t)$ is given by
\begin{equation}
H_0(t) =
\begin{pmatrix}
H_{\rm op}+E_{\rm 0}(t) & W_1 & W_2 \\
W_1 & H_{\rm cl_1}+E_{\rm 1}(t) & V \\
W_2 & V & H_{\rm cl_2}+E_{\rm 2}(t)
\end{pmatrix},
\label{eq:S_H0}
\end{equation}
where 
\begin{equation}
E_i(t) = \hbar \alpha_i\cos(\omega_{\rm I}t+\phi_{\rm I}) + \hbar \beta_i\cos(\omega_{\rm II}t+\phi_{\rm II}).
\label{amplitude}
\end{equation}
Here, \(i=0,1,2\); \(\alpha_i\) and \(\beta_i\) are the modulation amplitudes induced by the two modulation tones \(\omega_{\rm I}\) and \(\omega_{\rm II}\), respectively, and \(\phi_{\rm I}\) and \(\phi_{\rm II}\) are their phases.


We then apply the unitary transformation
\begin{equation}
U(t)= U_{00}(t)\ket{\rm op}\bra{\rm op}
+
U_{11}(t)\ket{\rm cl_1}\bra{\rm cl_1}
+
U_{22}(t)\ket{\rm cl_2}\bra{\rm cl_2},
\end{equation}
with 
\begin{eqnarray}
U_{00}(t)
&=&
e^{
-i\left[
\frac{\alpha_0}{\omega_{\rm I}}\sin(\omega_{\rm I}t+\phi_{\rm I}) +\frac{\beta_0}{\omega_{\rm II}}\sin(\omega_{\rm II}t+\phi_{\rm II})
\right]},
\nonumber\\
U_{11}(t)
&=&
e^{
-i\left[l\omega_{\rm II}t
+\frac{\alpha_1}{\omega_{\rm I}}\sin(\omega_{\rm I}t+\phi_{\rm I})
+\frac{\beta_1}{\omega_{\rm II}}\sin(\omega_{\rm II}t+\phi_{\rm II})
\right]},
\nonumber\\
U_{22}(t)
&=&
e^{
-i\left[
m\omega_{\rm I}t
+n\omega_{\rm II}t
+\frac{\alpha_2}{\omega_{\rm I}}\sin(\omega_{\rm I}t+\phi_{\rm I})
+\frac{\beta_2}{\omega_{\rm II}}\sin(\omega_{\rm II}t+\phi_{\rm II})
\right]},
\end{eqnarray}
where \(l\), \(m\), and \(n\) are integer Floquet indices.

The transformed Hamiltonian is
\begin{equation}
H_1(t)=U^\dagger(t)H_0(t)U(t)-i\hbar U^\dagger(t)\frac{\partial U(t)}{\partial t},
\end{equation}
written as
\begin{equation}
H_1(t)=
\begin{pmatrix}
H_{\rm op} & W_1 e^{-i\left[l\omega_{\rm II}t + \theta_{01}(t)\right]} & W_2e^{-i\left[m\omega_{\rm I}t+n\omega_{\rm II}t+\theta_{02}(t)\right]} \\
W_1^* e^{i\left[l\omega_{\rm II}t + \theta_{01}(t)\right]} & H_{\rm cl_1}- l \hbar \omega_{\rm II} & V e^{-i\left[ m\omega_{\rm I}t+(n-l)\omega_{\rm II}t + \theta_{12}(t)\right]} \\
W_2^*e^{i\left[m\omega_{\rm I}t+n\omega_{\rm II}t+\theta_{02}(t)\right]} & V^* e^{i\left[ m\omega_{\rm I}t+(n-l)\omega_{\rm II}t + \theta_{12}(t)\right]} & H_{\rm cl_2} -m\hbar \omega_{\rm I}-n\hbar \omega_{\rm II}
\end{pmatrix},
\label{eq:S_H1_time}
\end{equation}
where
\begin{eqnarray}
\theta_{01}
&=&
\frac{\alpha_1-\alpha_0}{\omega_{\rm I}}\sin(\omega_{\rm I}t+\phi_{\rm I})
+\frac{\beta_1-\beta_0}{\omega_{\rm II}}\sin(\omega_{\rm II}t+\phi_{\rm II}),
\nonumber\\
\theta_{02}
&=&
\frac{\alpha_2-\alpha_0}{\omega_{\rm I}}\sin(\omega_{\rm I}t+\phi_{\rm I})+
\frac{\beta_2-\beta_0}{\omega_{\rm II}}\sin(\omega_{\rm II}t+\phi_{\rm II}),
\nonumber\\
\theta_{12}
&=&
\frac{\alpha_2-\alpha_1}{\omega_{\rm I}}
\sin(\omega_{\rm I}t+\phi_{\rm I})
+\frac{\beta_2-\beta_1}{\omega_{\rm II}}\sin(\omega_{\rm II}t+\phi_{\rm II}).
\end{eqnarray}

Using the Jacobi--Anger expansion,
\begin{equation}
e^{ix\sin\theta}=\sum_{k=-\infty}^{\infty}J_k(x)e^{ik\theta},
\end{equation}
where $J_k(x)$ is the $k$-th order Bessel function of the first kind, and keeping the near-resonant terms ($E_1^{0}- l \hbar \omega_{\rm II}$ and $E_2^{0}-m\hbar\omega_{\rm I}-n\hbar\omega_{\rm II}$ are near resonance) under the rotating-wave approximation, the effective static Hamiltonian becomes
\begin{equation}
H_{\rm eff}=
\begin{pmatrix}
H_{\rm op} & W_1^{\rm eff} & W_2^{\rm eff} \\
\left(W_1^{\rm eff}\right)^* & H_{\rm cl_1} - l \hbar \omega_{\rm II} & V^{\rm eff} \\
\left(W_2^{\rm eff}\right)^* & \left(V^{\rm eff}\right)^* & H_{\rm cl_2}-m\hbar\omega_{\rm I}-n\hbar\omega_{\rm II}
\end{pmatrix}.
\label{eq:S_Heff}
\end{equation}
The effective couplings are
\begin{eqnarray}
W_1^{\rm eff}
&=&
W_1
J_0\left(\frac{\alpha_0-\alpha_1}{\omega_{\rm I}}\right)
J_l\left(\frac{\beta_0-\beta_1}{\omega_{\rm II}}\right)
e^{il\phi_{\rm II}},
\nonumber\\
W_2^{\rm eff}
&=&
W_2
J_m\left(\frac{\alpha_0-\alpha_2}{\omega_{\rm I}}\right)
J_n\left(\frac{\beta_0-\beta_2}{\omega_{\rm II}}\right)
e^{i(m\phi_{\rm I}+n\phi_{\rm II})},
\nonumber\\
V^{\rm eff}
&=&
V
J_m\left(\frac{\alpha_1-\alpha_2}{\omega_{\rm I}}\right) J_{n-l}\left(\frac{\beta_1-\beta_2}{\omega_{\rm II}}\right)
e^{i[m\phi_{\rm I}+(n-l)\phi_{\rm II}]}.
\end{eqnarray}

When \(\beta_0-\beta_1=0\), \(\beta_0-\beta_2=0\), 
\(\alpha_0-\alpha_2=0\), \(l=0\), \(n=0\), and \(m=-1\), we obtain an effective Hamiltonian where $W_2^{\rm eff} = 0$ and the other relevant couplings are renormalized by Bessel functions: the effective coupling between open channel and closed channel 1 is proportional to \(J_0(\frac{\alpha_0-\alpha_{1}}{\omega_{\rm I}})W_1\), while the effective coupling between the two closed channels is proportional to \(J_{-1}(\frac{\alpha_1-\alpha_2}{\omega_{\rm I}})V\). This is the ideal effective Hamiltonian for matter-wave induced transparency under single-frequency modulation.

When \(l=1\) and \(m=-1\), the effective Hamiltonian describes the dual-frequency engineering of matter-wave induced transparency in the main text, with the effective couplings controlled by the corresponding Bessel-function factors.

\subsection{Coupled Schrödinger Equations with the Effective Hamiltonian}
Next, we determine the scattering properties by solving a set of coupled stationary Schr\"{o}dinger equations using the effective Hamiltonian. Without loss of generality, we consider the ideal effective Hamiltonian for matter-wave induced transparency under single-frequency modulation. The corresponding stationary Schr\"{o}dinger equations with the state $\ket{\Psi}=\varphi_{\rm op}(\mathbf r)\ket{\rm op}+\varphi_{\rm cl_1}(\mathbf r)\ket{\rm cl_1}+\varphi_{\rm cl_2}(\mathbf r)\ket{\rm cl_2}$ are
\begin{equation}
\begin{cases}
H_{\rm op}\varphi_{\rm op}
+
W_1^{\rm eff}\varphi_{\rm cl_1}
=
E\varphi_{\rm op},
\\
W_1^{\rm eff}\varphi_{\rm op}
+
H_{\rm cl_1}\varphi_{\rm cl_1}
+
V^{\rm eff}\varphi_{\rm cl_2}
=
E\varphi_{\rm cl_1},
\\
V^{\rm eff}\varphi_{\rm cl_1}
+
\left(H_{\rm cl_2}+\hbar\omega_{\rm I}\right)\varphi_{\rm cl_2}
=
E\varphi_{\rm cl_2}.
\end{cases}
\label{eq:S_stationary}
\end{equation}
The last equation adds a single Floquet energy quantum \(\hbar\omega_{\rm I}\) to the second closed-channel Hamiltonian $H_{\rm cl_2}$. This term is essential for the resonance condition of the second molecular state \(\ket{m_2}\), which is typically off resonance without modulation.

At large distance, the open-channel wave function has the standard scattering form
\begin{equation}
\varphi_{\rm op}(\mathbf r)
\underset{r\rightarrow\infty}{\sim}
\frac{1}{(2\pi)^{3/2}}
\left[
e^{i\mathbf k\cdot\mathbf r}
+
f(\theta,k)\frac{e^{ikr}}{r}
\right],
\label{eq:S_asymptotic}
\end{equation}
where \(f(\theta,k)\) is the scattering amplitude. 
The Lippmann--Schwinger equations are
\begin{equation}
\begin{cases}
\ket{\varphi_{\rm op}}
=
\ket{\varphi_{\mathbf k}^{(+)}}
+
G_{\rm op}(E+i0^+)W_1^{\rm eff}\ket{\varphi_{\rm cl_1}},
\\
\ket{\varphi_{\rm cl_1}}
=
G_{\rm cl_1}(E)W_1^{\rm eff}\ket{\varphi_{\rm op}}
+
G_{\rm cl_1}(E)V^{\rm eff}\ket{\varphi_{\rm cl_2}},
\\
\ket{\varphi_{\rm cl_2}}
=
G_{\rm cl_2}(E)V^{\rm eff}\ket{\varphi_{\rm cl_1}},
\end{cases}
\label{eq:S_LS}
\end{equation}
with
\begin{equation}
\begin{cases}
G_{\rm op}(E+i0^+)=(E+i0^+-H_{\rm op})^{-1},
\\
G_{\rm cl_1}(E)=(E-H_{\rm cl_1})^{-1},
\\
G_{\rm cl_2}(E)=(E-H_{\rm cl_2}-\hbar\omega_{\rm I})^{-1}.
\end{cases}
\label{eq:S_Green}
\end{equation}
where $|\varphi_k^{(+)}\rangle$ is the background scattering state, which satisfies $H_{\rm op}|\varphi_k^{(+)}\rangle=\frac{\hbar^2k^2}{2\mu}|\varphi_k^{(+)}\rangle$. In the long-range approximation, the coordinate representation of $G_{\rm op}$ takes the form:

\begin{equation}
    G_{\rm op}(E,\mathbf{r},\mathbf{r'}) \sim -\frac{(2\pi)^{3/2}\mu}{2\pi \hbar^2}\frac{e^{ikr}}{r} [\varphi_{\mathbf{k}}^{(-)} (\mathbf{r'})]^{*},
    \label{G_function}
\end{equation}
where $\varphi_{\mathbf{k}}^{(-)} (\mathbf{r'}) = [\varphi_{\mathbf{-k}}^{(+)} (\mathbf{r'})]^*$. Near the two relevant molecular resonances, we keep only the resonant molecular states \(\ket{m_1}\) and \(\ket{m_2}\) in the two closed-channel Green's functions,
\begin{equation}
G_{\rm cl_1}(E)
\simeq
\frac{\ket{m_1}\bra{m_1}}{E-E_1^{\rm res}},
\qquad
G_{\rm cl_2}(E)
\simeq
\frac{\ket{m_2}\bra{m_2}}{E-E_2^{\rm res}}.
\label{eq:S_pole}
\end{equation}
Here, \(E_1^{\rm res}=E_1^{0}\), and \(E_2^{\rm res}=E_2^{0}+\hbar \omega_{\rm I}\).
Within this pole approximation, the resonant part of the open-channel scattering amplitude is governed by the dressed propagator of \(\ket{m_1}\),
\begin{equation}
D_1(E)
=
\frac{1}
{
E-\left[E_1^{\rm res} +
E_1^{\rm shift}(E)\right]
-
\frac{|V_{12}^{\rm eff}|^2}{E-E_2^{res}}
},
\label{eq:S_D1}
\end{equation}
where
\begin{equation}
V_{12}^{\rm eff}
=
\bra{m_1}V^{\rm eff}\ket{m_2},
\end{equation}
and
\begin{equation}
E_1^{\rm shift}(E)
=
\bra{m_1}W_1^{\rm eff} G_{\rm op}(E+i0^+) W_1^{\rm eff}\ket{m_1}
\end{equation}
is the energy shift arising from the coupling of \(\ket{m_1}\) to the continuum. 
The resonant scattering amplitude is therefore proportional to
\begin{equation}
f_{\rm res}(\theta,k)
\propto
-
\bra{\varphi_{\mathbf k}^{(-)}}W_1^{\rm eff}\ket{m_1}
D_1(E)
\bra{m_1}W_1^{\rm eff}\ket{\varphi_{\mathbf k}^{(+)}},
\label{eq:S_fres}
\end{equation}
and the total scattering amplitude is \(f=f_{bg}+f_{\rm res}\). 
The resonant state that couples directly to the open channel in Eq.~\eqref{eq:S_fres} is \(\ket{m_1}\), because \(W_1\) couples \(\ket{\rm op}\) to \(\ket{\rm cl_1}\).

In the low-energy \(s\)-wave limit, the corresponding scattering length can be written in the effective form
\begin{equation}
a
= -f =
a_{bg}
\left[
1-
\frac{\Delta_B^{\rm eff}}
{B-B_0-\Delta_{12}^{\rm eff}/\delta(B,\omega_{\rm I})}
\right],
\label{eq:S_a_lossless}
\end{equation}
where \(a_{bg}\) is the background scattering length, \(B_0\) is the shifted position of the Feshbach resonance associated with \(\ket{m_1}\), and
\begin{equation}
\delta(B,\omega_{\rm I})
=
\omega_{\rm I}-\omega_2^0(B)
\end{equation}
is the two-photon-like detuning between \(\ket{m_2}\) and the atomic scattering state in the Floquet frame. 
Here
\begin{equation}
\omega_2^0(B)
=
\frac{-E_2^0(B)}{\hbar}
\end{equation}
is the frequency needed to bring \(\ket{m_2}\) into resonance with the open-channel threshold in the chosen convention. 
The parameters \(\Delta_B^{\rm eff}\) and \(\Delta_{12}^{\rm eff}\) characterize, the strength of the \(\ket{m_1}\) Feshbach coupling and the modulation-assisted coupling between the two closed-channel molecular states, respectively. 
Up to convention-dependent signs, they are proportional to
\begin{equation}
\Delta_B^{\rm eff}
\propto
J_0^2(\frac{\alpha_0-\alpha_1}{\omega_{\rm I}})
\left|
\bra{m_1}W_1\ket{\varphi_{\mathbf 0}^{(+)}}
\right|^2,
\end{equation}
and
\begin{equation}
\Delta_{12}^{\rm eff}
\propto
\left|
J_1(\frac{\alpha_1-\alpha_2}{\omega_{\rm I}})
\bra{m_1}V\ket{m_2}
\right|^2.
\end{equation}
Equation~\eqref{eq:S_a_lossless} shows the origin of matter-wave induced transparency. 
At the two-photon-like resonance,
\begin{equation}
\delta(B,\omega_{\rm I})=\omega_{\rm I}-\omega_2^0(B)=0,
\end{equation}
the dressing term \(\Delta_{12}^{\rm eff}/\delta\) becomes singular in the ideal lossless limit. 
Consequently, the resonant contribution to the scattering amplitude vanishes and the scattering length approaches the background value,
\begin{equation}
a\rightarrow a_{bg}.
\end{equation}
Thus, the otherwise strong Feshbach scattering is suppressed by destructive interference through the second molecular state.

\subsection{Phenomenological Treatment of Inelastic Collisions}
In our treatment of the modulated three-channel model presented above, the inelastic process of scattering into higher Floquet modes is ignored within the rotating-wave approximation~\cite{Ref1}. Beyond this process, a real experimental system can involve other inelastic mechanisms, such as spin relaxation into other open channels not included in the three-channel model \cite{Ref3,RevModPhys.82.1225,Smith2015}, as well as optical- or RF-induced bound-free or bound-bound transitions \cite{Ref4,Ref5,Ref6}. To include these inelastic processes, we introduce decay phenomenologically into the two closed-channel molecular states, with the complex scattering length given by
\begin{equation}
a
=
a_{bg}
\left[
1-
\frac{\Delta^{\rm eff}_B}
{
(B-B_0)
-
\frac{\Delta_{12}^{\rm eff}}{\delta(B,\omega_{\rm I})-i\frac{\gamma_2}{2}}
-
i\frac{\gamma_1}{2}\frac{\hbar}{\Delta\mu_1}
}
\right].
\label{eq:S_a_complex}
\end{equation}
Here \(\gamma_1\) and \(\gamma_2\) are the decay rates of \(\ket{m_1}\) and \(\ket{m_2}\) in angular-frequency units, respectively, and $\Delta\mu_1$ is the magnetic moment difference between \(\ket{m_1}\) and the free-atom state.
At the transparency condition \(\delta=0\), Eq.~\eqref{eq:S_a_complex} gives
\begin{equation}
a(\delta=0)
=
a_{bg}
\left[
1
-
\frac{i\Delta_B^{\rm eff}\gamma_2/2}
{\Delta_{12}^{\rm eff}+\frac{\hbar\gamma_1\gamma_2}{4\Delta\mu_1}+i\frac{\gamma_2}{2}(B-B_0)}
\right].
\label{eq:S_a_transparency}
\end{equation}
This expression shows that the imaginary part of the scattering length, and therefore the inelastic loss rate, is suppressed by the closed-channel coupling scale \(\Delta_{12}^{\rm eff}\). In particular, when \(\gamma_2=0\), the inelastic loss rate is completely suppressed. Thus, efficient loss suppression requires a sufficiently strong modulation-assisted coupling between \(\ket{m_1}\) and \(\ket{m_2}\), and a relatively small decay rate \(\gamma_2\) of the second molecular state.

\subsection{Beyond the Ideal Effective Hamiltonian for Matter-wave Induced Transparency}
Now we consider the general case of Eq.\ref{eq:S_Heff} where the effective coupling $W_2^{\rm eff}$ is nonzero. The same procedure for solving the coupled Schrödinger equations can be carried out straightforwardly as in the ideal case, yielding the corresponding scattering length as
\begin{equation}
a
=a_{bg}\left[1 - \frac{\Gamma_{10}^{\rm eff}\left(E_2^{\rm res}+E_2^{\rm shift}\right)+\Gamma_{20}^{\rm eff}\left(E_1^{\rm res}+E_1^{\rm shift}\right)-2(V_{12}^{\rm eff}+G_{12}^{\rm eff})w_1^{\rm eff}w_2^{\rm eff}}{\left(E_1^{\rm res}+E_1^{\rm shift}\right)\left(E_2^{\rm res}+E_2^{\rm shift}\right)-(V_{12}^{\rm eff}+G_{12}^{\rm eff})^2} \right],
\end{equation}
where,
\begin{eqnarray}
E_1^{\rm res}
&=&
E_1^0- l \hbar \omega_{\rm II},
\nonumber\\
E_2^{\rm res}
&=&
E_2^0-m\hbar\omega_{\rm I}-n\hbar\omega_{\rm II},
\end{eqnarray}
\begin{eqnarray}
E_1^{\rm shift}
&=&
\bra{m_1}W_1^{\rm eff} G_{\rm op}(i0^+) W_1^{\rm eff}\ket{m_1},
\nonumber\\
E_2^{\rm shift}
&=&
\bra{m_2}W_2^{\rm eff} G_{\rm op}(i0^+) W_2^{\rm eff}\ket{m_2}.
\end{eqnarray}
The resonance strengths $\Gamma_{10}^{\rm eff}$ and $\Gamma_{20}^{\rm eff}$ characterize the coupling of molecular states \(\ket{m_1}\) and \(\ket{m_2}\) to the atomic scattering state, respectively. These quantities are equal to
\begin{eqnarray}
\Gamma_{10}^{\rm eff}
&=&
\left(w_1^{\rm eff}\right)^2,
\nonumber\\
\Gamma_{20}^{\rm eff}
&=&
\left(w_2^{\rm eff}\right)^2,
\end{eqnarray}
and
\begin{eqnarray}
w_1^{\rm eff}
&\propto&
\bra{m_1}W_1^{\rm eff}\ket{\varphi_{\mathbf 0}^{(+)}},
\nonumber\\
w_2^{\rm eff}
&\propto&
\bra{m_2}W_2^{\rm eff}\ket{\varphi_{\mathbf 0}^{(+)}}.
\end{eqnarray}

$V_{12}^{\rm eff}$ results from the coupling between the two closed channel molecular states, while $G_{12}^{\rm eff}$ is due to the simultaneous coupling between the two closed channels and the open channel. Specifically,
\begin{equation}
V_{12}^{\rm eff}
=
\bra{m_1}V^{\rm eff}\ket{m_2},
\end{equation}
and
\begin{equation}
G_{12}^{\rm eff}
\propto
\bra{m_1}W_1^{\rm eff} G_{\rm op}(i0^+) W_2^{\rm eff}\ket{m_2}.
\end{equation}

Inelastic collisions are included via the same phenomenological process as before, where the complex scattering length is expressed as
\begin{equation}
a
=a_{bg}\left[1 - \frac{\Gamma_{10}^{\rm eff}\left(\delta-i\hbar\gamma_2/2\right)+\Gamma_{20}^{\rm eff}\left(\Delta-i\hbar\gamma_1/2\right)-2(V_{12}^{\rm eff}+G_{12}^{\rm eff})w_1^{\rm eff}w_2^{\rm eff}}{\left(\Delta-i\hbar\gamma_1/2\right)\left(\delta-i\hbar\gamma_2/2\right)-(V_{12}^{\rm eff}+G_{12}^{\rm eff})^2} \right],
\label{eq36}
\end{equation}
where $\delta$ and $\Delta$ are defined as
\begin{equation}
    \delta= E_2^{\rm res}+E_2^{\rm shift},
\end{equation}
\begin{equation}
    \Delta = E_1^{\rm res}+E_1^{\rm shift},
\end{equation}
and the imaginary part of this expression is guaranteed to be non-positive.
In the next section, this method of manually adding an imaginary decay rate component to the resonance energy term is shown to be consistent with the results from multichannel quantum defect theory.

Based on Eq.~\ref{eq36}, we reexamine the dark state condition ($\mathrm{Im}(a)=0$) when $W_2^{\rm eff}$ is non-zero and $\gamma_2=0$. Straightforward algebra shows that the dark state condition shifts from zero to $\frac{\left(V_{12}^{\rm eff}+G_{12}^{\rm eff}\right)w_2^{\rm eff}}{w_1^{\rm eff}}$.

\subsection{Fano Line Shape of the Complex Scattering Length}
Equation~\ref{eq36} exhibits Fano profiles when we scan $\Delta=E_1^{\rm res}+E_1^{\rm shift}$ and $\delta=E_2^{\rm res}+E_2^{\rm shift}$ as linear functions of a real-valued variable $\varepsilon$ (i.e., $\Delta=b_1+a_1\varepsilon$ and $\delta=b_2+a_2\varepsilon$). We demonstrate this as follows.

First, Eq.~\ref{eq36} can be written as
\begin{equation}
    a = a_{bg} \left[ p + \frac{s_1}{\varepsilon-z_1} + \frac{s_2}{\varepsilon-z_2}\right],
\end{equation}
where $p$, $s_1$, and $s_2$ are complex constants and $z_1$ and $z_2$ are complex roots. The real part and the imaginary part of $\frac{s_k}{\varepsilon-z_k}$ each take the form
\begin{equation}
    \frac{A_k^{r/i}+B_k^{r/i}\varepsilon}{(\varepsilon-\alpha_k)^2+\beta_k^2},
\label{fano1}
\end{equation}
where real-valued $\alpha_k$ and $\beta_k$ depend on the $k$th root $z_k$, and real-valued $A_k^{r/i}$ and $B_k^{r/i}$ depend on the $k$th root $z_k$, as well as on the real and imaginary parts of $s_k$. Then, Eq.~\ref{fano1} can be expressed as a Fano profile \cite{Fano}
\begin{equation}
    c_k\frac{(q_k+\epsilon_k)^2}{1+\epsilon_k^2}-c_k,
\end{equation}
where $\epsilon_k=(\varepsilon-\alpha_k)/\left|\beta_k\right|$, and $c_k$ and $q_k$ are determined from Eq.~\ref{fano1}.

Therefore, both the real part of the scattering length and the inelastic loss rate (or the imaginary part of the scattering length, $K_{inel} = -\frac{2h}{\mu}\mathrm{Im}(a)$ for BEC) are sums of several Fano line shapes.

\subsection{A Minimal Non-Hermitian Three-Level Model}
As in the main text, the same physics can be summarized by a minimal non-Hermitian three-level Hamiltonian
\begin{equation}
H_{\rm min}
=
\hbar
\begin{pmatrix}
0 & \Omega_{01}^{\rm eff}/2 & \Omega_{02}^{\rm eff}/2 \\
(\Omega_{01}^{\rm eff})^{*}/2 & \Delta-i\gamma_1/2 & \Omega_{12}^{\rm eff}/2 \\
(\Omega_{02}^{\rm eff})^{*}/2 & (\Omega_{12}^{\rm eff})^{*}/2 & \delta-i\gamma_2/2
\end{pmatrix}.
\label{eq:S_Hmin}
\end{equation}
Here, \(\Omega_{01}^{\rm eff}\) is the effective coupling between the free-atom state \(\ket{a}\) and \(\ket{m_1}\), \(\Omega_{12}^{\rm eff}\) is the modulation-assisted coupling between the two closed channel molecular states, and \(\Omega_{02}^{\rm eff}\) is a weak residual direct coupling between \(\ket{a}\) and \(\ket{m_2}\). 
In the ideal limit \(\Omega_{02}^{\rm eff}=0\), \(\delta=0\) and $\gamma_2=0$, the system supports a dark state that avoids the lossy intermediate molecular state \(\ket{m_1}\),
\begin{equation}
\ket{D}
\propto
\Omega_{12}^{\rm eff}\ket{a}
-
\Omega_{01}^{\rm eff}\ket{m_2}.
\end{equation}
This dark-state structure is the origin of matter-wave induced transparency.
A small but nonzero \(\Omega_{02}^{\rm eff}\) will slightly shift the dark state condition of $\delta=0$ and produce an asymmetric line shape. A nonzero $\gamma_2$ will turn the dark state into a gray state due to the decay of $\ket{m_2}$, which requires a stronger $\Omega_{12}^{\rm eff}$ to reduce this effect.

\section{QUANTUM DEFECT TREATMENT}
\label{section2}

In the previous section, inelastic loss was incorporated phenomenologically by adding imaginary decay terms to the molecular resonance energies. This treatment provides a compact and physically transparent description of loss suppression in matter-wave induced transparency. However, in a real multichannel scattering problem, inelastic loss arises from coupling to additional open channels rather than from an intrinsically complex Hamiltonian. To make this connection explicit, we use multichannel quantum defect theory (MQDT) to describe the loss channels on the same footing as the elastic scattering channel. In this framework, the imaginary part of the complex scattering length emerges from the unitary scattering matrix after eliminating the closed channels and the additional inelastic open channels. The MQDT treatment therefore provides a microscopic justification for the phenomenological complex-energy model used above.

\subsection{General Description of Multichannel Quantum Defect Theory}
To represent the inelastic scattering properties of the modulated three-channel model more analytically, we apply multichannel quantum defect theory (MQDT) to this problem. 
In MQDT theory~\cite{Ref2}, we begin with a real-valued reaction matrix $K^{red}$, defined by relating the long-range behavior of the radial wave function $F(R)$ in the open channels to standing waves at large $R$:

\begin{equation}
    F_{ii'}\sim\sqrt{\frac{2\mu}{\pi \hbar^2k_i}}[\sin k_iR\delta_{ii'}+\cos k_iRK^{red}_{ii'}], \ \ R\rightarrow \infty
\end{equation}
where $F=\{F_{ii'}\}$ is a matrix of solutions of coupled-channel Schr\"{o}dinger equation, with the index $i$ denoting the channel, $i'$ denoting linearly independent solutions, $k_i$ the wave number in channel $i$, and $\mu$ the reduced mass. 

This matrix $K^{red}$ can be reduced from a matrix $K$ for open and closed channels, with the specific expression given by
\begin{equation}
    K^{red} =K^{OO}-K^{OC}(\tan\nu+K^{CC})^{-1}K^{CO},
\label{EqKred}
\end{equation}
where $K$ is divided into open and closed blocks as
\begin{equation}
    K =
    \begin{pmatrix}
    K^{OO} & K^{OC} \\
    K^{CO} & K^{CC} \\
    \end{pmatrix}.    
\end{equation}
Here $\nu_i(E)$ stands for a quantum defect phase that describes the phase
accumulation of bound-state levels in a molecular potential relative to a
reference energy. Near a closed-channel resonance, where $\nu_k\sim n\pi$, we use the approximation $\tan\nu_k\sim\frac{\partial\nu_k}{\partial E}(E-E_k^{res}-E_k^{shift})$.


The physical scattering matrix $S$ is then obtained as
\begin{equation}
    S=\exp(i\eta) \left(1+iK^{\rm red}\right)
    \left(1-iK^{\rm red}\right)^{-1}\exp(i\eta),
\label{Smatrix}
\end{equation}
where $\exp(i\eta)$ a diagonal matrix whose diagonal term $\eta_i$ represents the elastic phase shift for open channel $i$.

\subsection{Application of MQDT to Matter-wave Induced Transparency}
In Section~\ref{three-channel}, we include inelastic processes by directly adding an imaginary decay rate component to the corresponding resonance energy term. Here, these are treated by introducing an artificial open channel coupled to the corresponding closed decay channel. 

We start with a model with five channels: an incident open channel labeled 0, two closed molecular channels labeled $b_1$ and $b_2$ that support the molecular states $\ket{m_1}$ and $\ket{m_2}$, respectively, and two artificial open channels representing the dissipation of $b_1$ and $b_2$, labeled $a_1$ and $a_2$. Following the same approximations and procedures as in Ref.~\cite{Ref2}, the K matrix for this model is

\begin{equation}
    K = \begin{pmatrix}
    0 & 0 & 0 & K_{0b_1} & K_{0b_2}\\
    0 & 0 & 0 & K_{a_1b_1} & 0\\
    0 & 0 & 0 & 0 & K_{a_2b_2}\\
    K_{b_10} & K_{b_1a_1} & 0 & K_{b_1b_1} & K_{b_1b_2}\\
    K_{b_20} & 0 & K_{b_2a_2} & K_{b_2b_1}& K_{b_2b_2},\\
    \end{pmatrix}  
\end{equation}
where $K_{0b_i}$ is proportional to the coupling strength between the open channel 0 and the closed channel $b_i$, $K_{a_ib_i}$ describes inelastic scattering from the closed channel $b_i$ to the artificial open channel $a_i$, $K_{b_ib_i}$ is associated with the resonance energy shift from the coupling between the open channel 0 and the closed channel $b_i$, and $K_{b_1b_2}$ is a consequence of the coupling between the two closed channels and their mutual coupling to the open channel.

According to Eq.~\ref{EqKred} and using a similar notation as in Ref.~\cite{Ref2}, the corresponding reduced matrix $K$ reads:
\begin{equation}
    K^{red} = -\frac{1}{2}\frac{1}{d_1d_2-g^2}
    \begin{pmatrix}
    d_2\Gamma_1+d_1\Gamma_2-2g\sqrt{\Gamma_1\Gamma_2} & d_2\sqrt{\Gamma_1\gamma_1}-g\sqrt{\Gamma_2\gamma_1} & -g\sqrt{\Gamma_1\gamma_2}+d_1\sqrt{\Gamma_2\gamma_2} \\
    d_2\sqrt{\Gamma_1\gamma_1}-g\sqrt{\Gamma_2\gamma_1} & d_2\gamma_1 & -g\sqrt{\gamma_1\gamma_2} \\
    -g\sqrt{\Gamma_1\gamma_2}+d_1\sqrt{\Gamma_2\gamma_2} & -g\sqrt{\gamma_1\gamma_2} & d_1\gamma_2 \\
    \end{pmatrix} 
\end{equation}
where $d_1=E- \left(E_1^{\rm res}+E_1^{\rm shift}\right)$, $d_2=E- \left(E_2^{\rm res}+E_2^{\rm shift}\right)$, $-g = V_{12}^{\rm eff}+G_{12}^{\rm eff}$, $\Gamma_1 \propto
\Gamma_{10}^{\rm eff}$, $\Gamma_2 \propto
\Gamma_{20}^{\rm eff}$. $\gamma_1$ and $\gamma_2$ are the decay rates of the two molecular states, respectively.

Then we can deduce the incident channel component $S_{00}$ of the scattering matrix $S$ according to Eq.~\ref{Smatrix} and calculate the real and imaginary parts of the scattering length in the limit of the incident wave number $k\rightarrow0$, as given by
\begin{equation}
    a_r = a_{bg}-\frac{1}{2k}\frac{\rm Im(S_{00})}{\rm Re(S_{00})},
\end{equation}
\begin{equation}
    a_{i}=-\frac{1}{4 k}\left(1- \left|S_{00}\right|^2\right);
\end{equation}


\begin{figure*}[hbtp]
    \centering
    \includegraphics[width=1.0\textwidth]{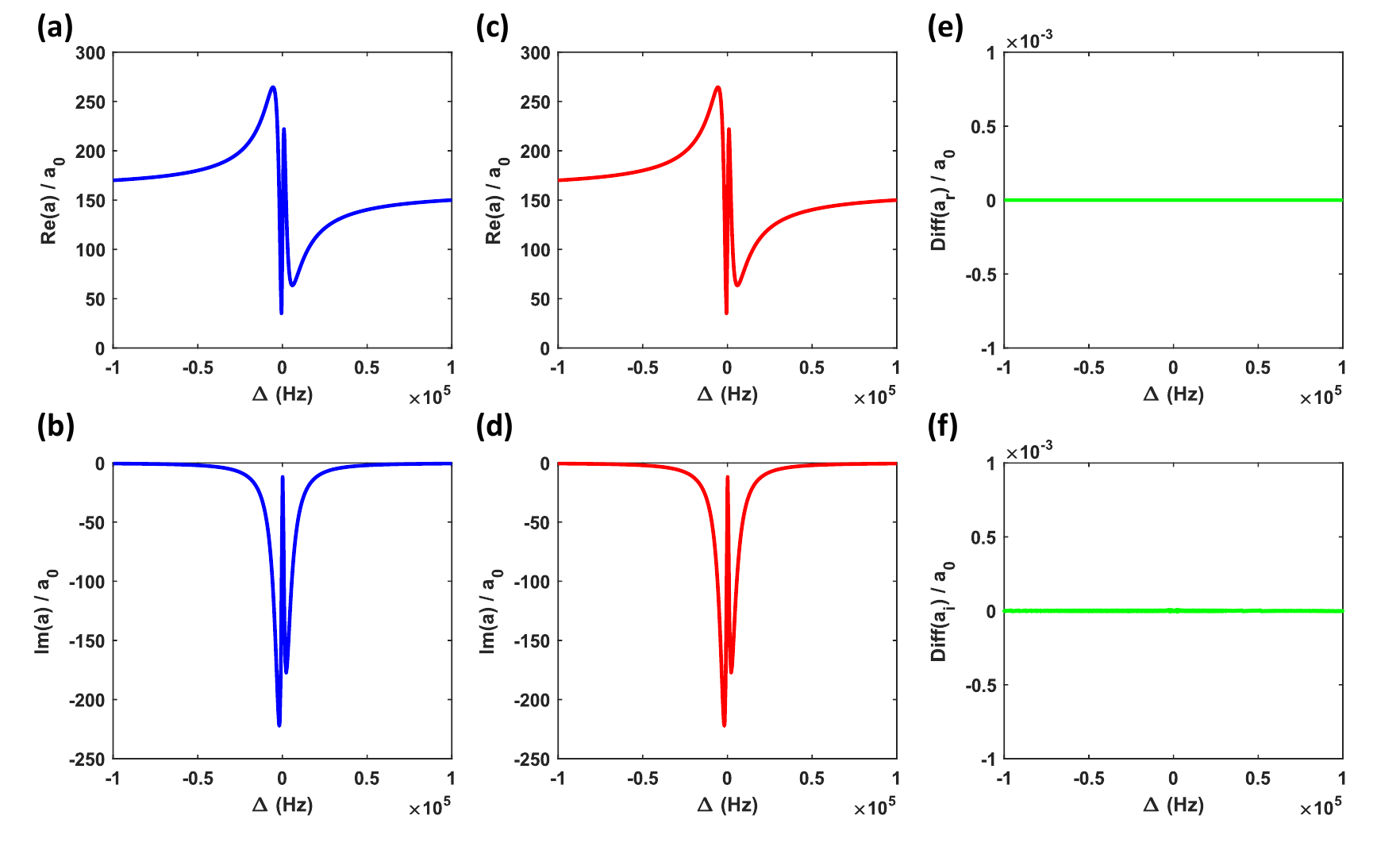}
    \caption{\textcolor{black}{Complex scattering length calculated with MQDT and Eq.~\ref{eq36}. Panels (a) and (b) show the real and imaginary parts of the scattering length obtained from MQDT, with parameters $\Delta = -d_1$, $\delta=-d_2=\Delta$, $E/h=10^{-12}$ Hz, $a_{bg}=160 a_0$, $g=-2$ kHz, $\gamma_1=10$ kHz, $\gamma_2=0.1$ kHz, $\Gamma_1 = 2ka_{bg}\Gamma_{10}^{\rm eff}$, $\Gamma_2 = 2ka_{bg}\Gamma_{20}^{\rm eff}$, $\Gamma_{10}^{\rm eff}=6.255$ kHz, and $\Gamma_{20}^{\rm eff}=0.021$ kHz. Panels (c) and (d) show the corresponding results from Eq.~\ref{eq36}, with $w_1^{\rm eff}=\sqrt{\Gamma_{10}^{\rm eff}}$, $w_2^{\rm eff}=\sqrt{\Gamma_{20}^{\rm eff}}$, $V_{12}^{\rm eff}+G_{12}^{\rm eff}=2$ kHz, and all other parameters the same as in the MQDT calculation. Panels (e) and (f) show the differences between the MQDT results and those from Eq.~\ref{eq36} for $\rm Re(a)$ and $\rm Im(a)$, respectively. These two approaches yield the same values within the computational error.}}  
    \label{com}
\end{figure*}

\subsection{Comparison to Section I Complex Scattering Length Expression} 
Here, we compare the complex scattering length obtained from MQDT with the expression in Eq.~\ref{eq36}. Figure~\ref{com} shows such an example, where panels (a) and (b) are the real and imaginary parts of the scattering length obtained from MQDT, panels (c) and (d) are the corresponding results from Eq.~\ref{eq36}, and panels (e) and (f) show the differences between the MQDT results and the results of Eq.~\ref{eq36} for $\rm Re(a)$ and $\rm Im(a)$, respectively. These two approaches yield the same values within the computational error.


\section{EFFECTS OF VARIOUS CONFIGURATIONS}
\label{section3}
In this section, we analyze how different effective level configurations modify the scattering response of matter-wave induced transparency. The ideal configuration provides a direct matter-wave analogue of optical EIT, in which destructive interference suppresses the resonant inelastic loss and restores the scattering length to its background value. Under realistic experimental conditions, however, a weak residual coupling between the open channel and the second molecular state can be present, and the two molecular resonances can interfere with each other. These effects shift the dark-state condition, generate asymmetric Fano line shapes, and give rise to bound states in the continuum (BICs). Therefore, the discussion below connects the simplified three-level picture to the experimentally observed line shapes and clarifies the physical origin of the BIC signatures observed in our measurements.

\subsection{Ideal Level Configuration and Matter-Wave Correspondence}
We start with the ideal level configuration, where the coupling strength between the molecular state $\ket{m_2}$ and the free-atom state $\ket{a}$ is zero (i.e., $W_2^{\rm eff}$ is zero). 

When we fix $\Delta=0$ and scan $\delta$, both real and imaginary parts of the scattering length exhibit a single Fano line shape, with the width (defined as in the Methods) given by $\left(2g \right)^2/\gamma_1+\gamma_2$. The maximum transparency of the imaginary part of the scattering length (i.e., the inelastic loss rate) is located at $\delta=0$, where the enhancement of the real part also vanishes.

\begin{figure*}[hbtp]
    \centering
    \includegraphics[width=0.9\textwidth]{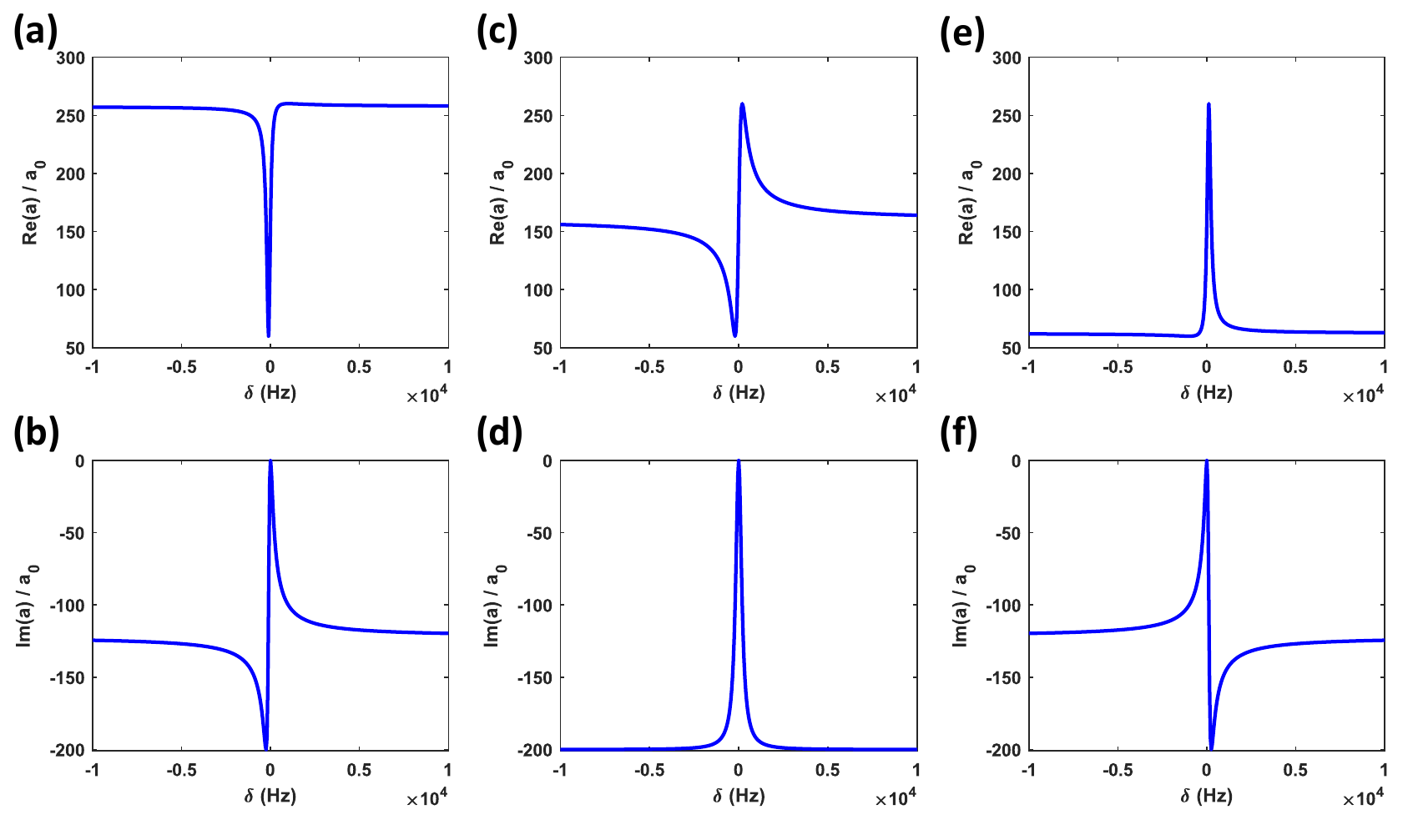}
    \caption{\textcolor{black}{Complex scattering length as a function of $\delta$ for fixed $\Delta$. Here, $E/h=10^{-12}$ Hz, $a_{bg}=160$ $a_0$, $g=-1$ kHz, $\gamma_1=10$ kHz, $\gamma_2=0$ kHz, $\Gamma_{10}^{\rm eff}=6.255$ kHz, and $\Gamma_{20}^{\rm eff}=0$ kHz. The values of $\Delta$ are -4 kHz, 0 kHz, 4 kHz from left to right panels, respectively. For these values, at the dark state condition $\delta=0$, the inelastic loss rate is completely suppressed and the enhancement of the real part of the scattering length vanishes as well. The widths in (c) and (d) are 0.400 kHz, which is equal to $\left(2g \right)^2/\gamma_1+\gamma_2$. The widths in (a), (b), (e), and (f) decrease to 0.244 kHz as the detuning increases. The absolute maxima in (b) and (f) are located at $\pm0.25$ kHz, respectively, which equal $g^2/\Delta$ and correspond to one of the diagonalized states in the coupled two-level system formed by the bare molecular states $\ket{m_1}$ and $\ket{m_2}$.}}  
    \label{sup2}
\end{figure*}

When we fix $\delta-\Delta=0$ while scanning $\Delta$, the profiles are sums of two Fano line shapes, with maximum transparency at $\Delta=\delta=0$. Here, the complex scattering length is a matter wave analogue of the complex susceptibility in EIT, where light absorption corresponds to inelastic loss rate $K_{inel} = -\frac{2h}{\mu}\mathrm{Im}(a)$ in matter waves, while the refractive index of the medium corresponds to nonlinear matter wave interactions $\mathrm{Re}(a)$. This correspondence may inspire applications for matter waves analogous to those in EIT, such as controlling the group velocity of colliding BECs \cite{Group}. 

\begin{figure*}[hbtp]
    \centering
    \includegraphics[width=0.9\textwidth]{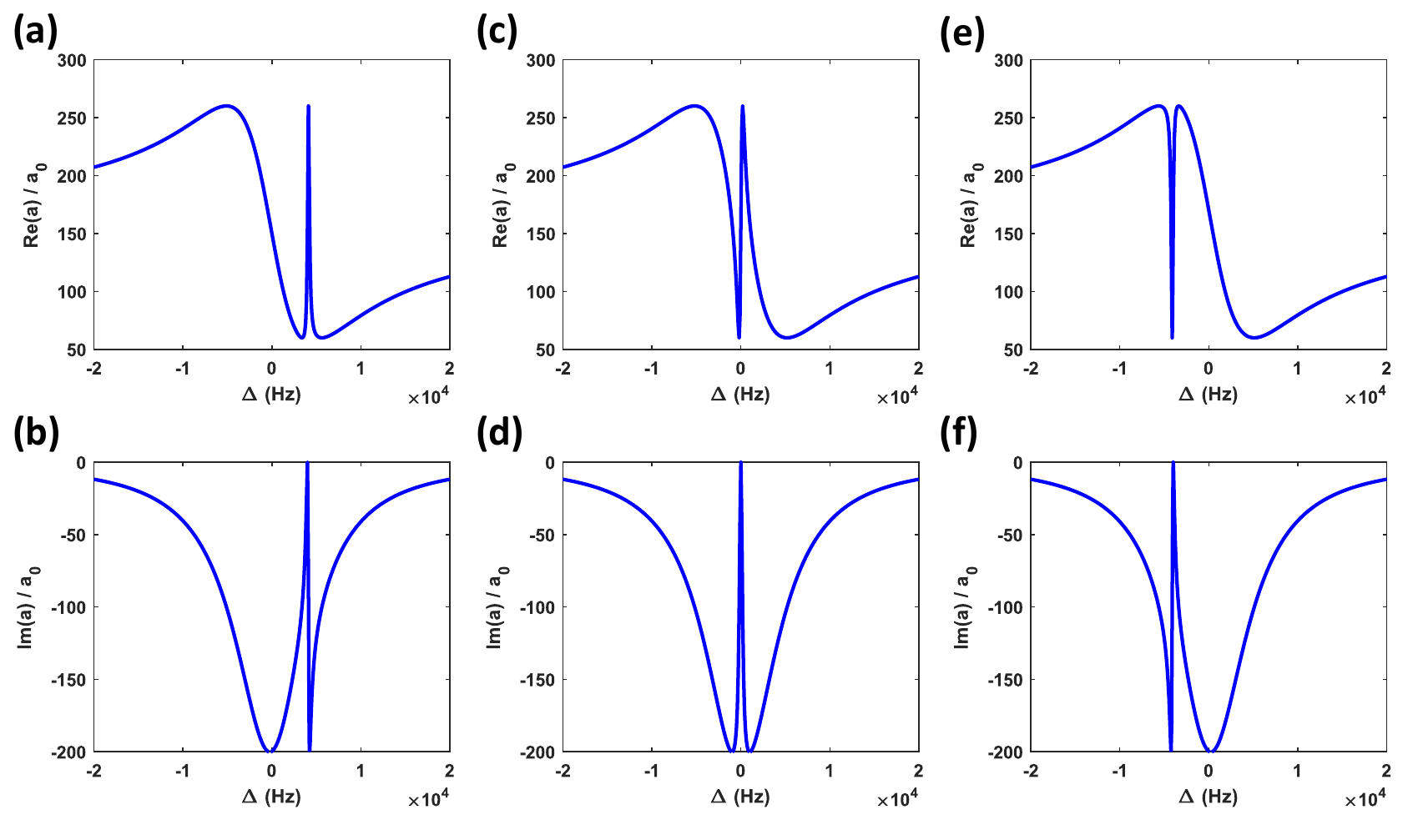}
    \caption{\textcolor{black}{Complex scattering length as a function of $\Delta$ for fixed $\delta -\Delta$. Here, $E/h=10^{-12}$ Hz, $a_{bg}=160$ $a_0$, $g=-1$ kHz, $\gamma_1=10$ kHz, $\gamma_2=0$ kHz, $\Gamma_{10}^{\rm eff}=6.255$ kHz, and $\Gamma_{20}^{\rm eff}=0$ kHz. The values of $\delta-\Delta$ are -4 kHz, 0 kHz, 4 kHz from left to right panels, respectively. The dark state condition $\delta=0$ completely suppresses the inelastic loss rate and eliminates the enhancement of the real part of the scattering length. Here, the complex scattering length is a matter wave analogue of the complex susceptibility in EIT, where light absorption corresponds to inelastic loss rate $K_{inel} = -\frac{2h}{\mu}\mathrm{Im}(a)$ in matter waves, while the refractive index of the medium corresponds to nonlinear matter wave interactions $\mathrm{Re}(a)$. The widths in (c) and (d) are 0.417 kHz, which is close to $\left(2g \right)^2/\gamma_1+\gamma_2=0.400$ kHz. The widths in (a), (b), (e), and (f) decrease to 0.240 kHz as the detuning increases. The absolute local maxima in (d) are located at $\pm1.00$ kHz. Those in (b) are at $-0.236$ kHz and $4.236$ kHz, while those in (f) are at $-4.236$ kHz and $0.236$ kHz. These values equal $\frac{-(\delta-\Delta)\pm \sqrt{\left(\delta-\Delta \right)^2+4g^2}}{2}$.}}  
    \label{sup3}
    \vfill
    \centering
    \includegraphics[width=0.9\textwidth]{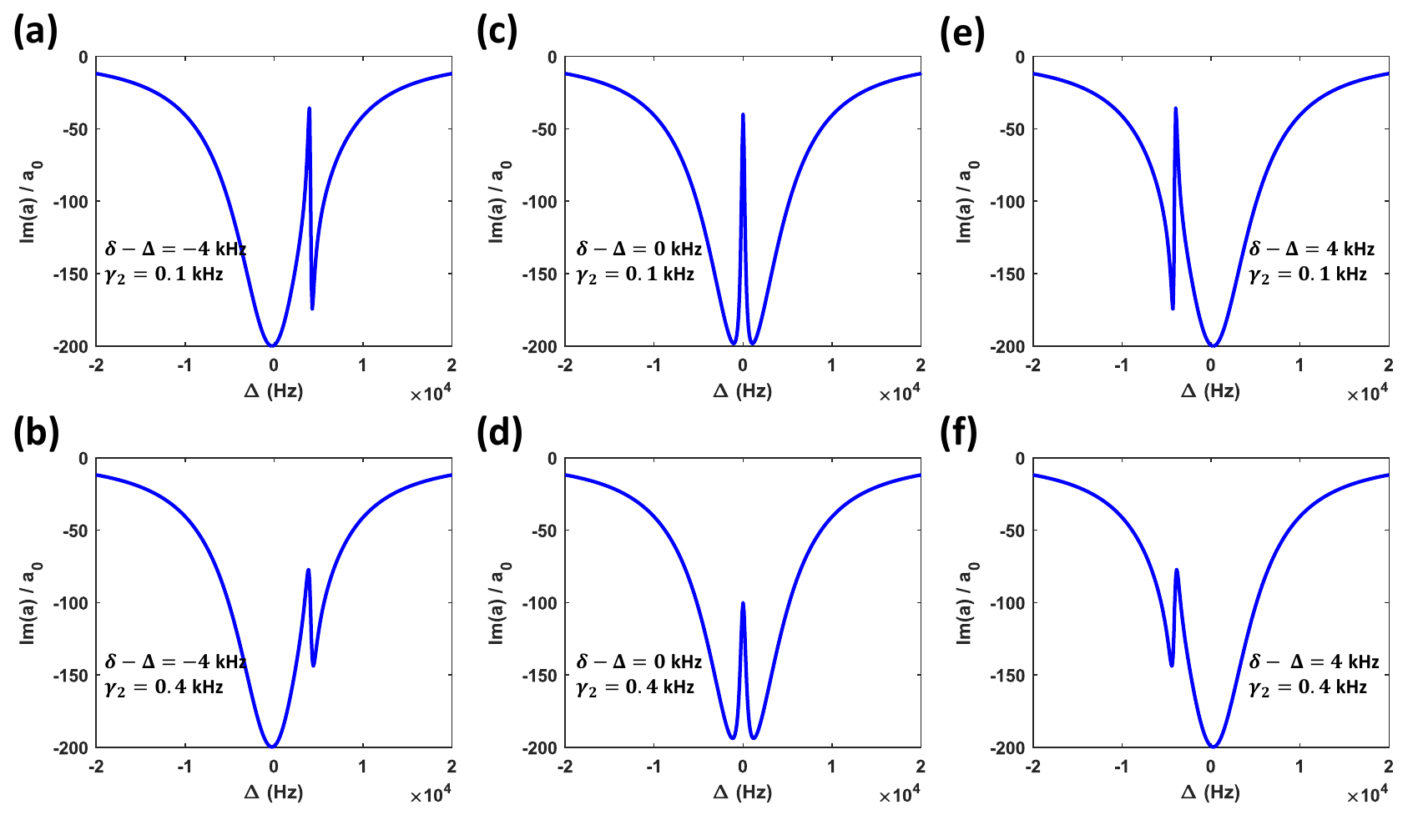}
    \caption{\textcolor{black}{Imaginary part of the scattering length as a function of $\Delta$ for fixed $\delta -\Delta$. Here, $E/h=10^{-12}$ Hz, $a_{bg}=160$ $a_0$, $g=-1$ kHz, $\gamma_1=10$ kHz, $\Gamma_{10}^{\rm eff}=6.255$ kHz, and $\Gamma_{20}^{\rm eff}=0$ kHz. The values of $\delta-\Delta$ are -4 kHz, -4 kHz, 0 kHz, 0 kHz, 4 kHz, 4 kHz for (a)-(f), respectively. The values of $\gamma_2$ are 0.1 kHz, 0.4 kHz, 0.1 kHz, 0.4 kHz, 0.1 kHz, 0.4 kHz for (a)-(f), respectively. The height of the transparency window decreases as $\gamma_2$ increase. The positions of absolute local minima are 3.961 kHz, 3.860 kHz, 0.000 kHz, 0.000 kHz, -3.961 kHz, -3.860 kHz for (a)-(f), respectively. This shows that, except at zero detuning, a nonzero $\gamma_2$ gradually shifts the transparency condition away from $\delta=0$. The narrow Fano profile widths in (c) and (d) are 0.522 kHz and 0.837 kHz, respectively, close to $\left(2g \right)^2/\gamma_1+\gamma_2$ 0.500 kHz and 0.800 kHz. The widths in (a), (b), (e), and (f) decrease to 0.341 kHz, 0.642 kHz, 0.341 kHz, and 0.642 kHz as the detuning increases.}}  
    \label{sup4}    
\end{figure*}

The broad Fano profile of $\mathrm{Im}(a)$ characterizes the Feshbach loss associated with $\ket{m_1}$, and the narrow profile quantifies the transparency window induced by the coupling between $\ket{m_1}$ and $\ket{m_2}$. Similar to the case of scanning $\delta$, the width of the narrow Fano profile is $\left(2g \right)^2/\gamma_1+\gamma_2$ when $\left|g\right| \ll \gamma_1$. When $\gamma_2=0$, the inelastic loss rate can be completely suppressed and the distance between the two loss maxima is $2\left|g\right|$. A nonzero $\gamma_2$ transfers the ideal dark state to a less-dark state, requiring $\left(2g \right)^2/\gamma_1 \gg \gamma_2$ to fully suppress the inelastic loss rate. At the same time, the distance between the two loss maxima increases compared to $2\left|g\right|$.

When we scan $\delta$ with a fixed nonzero $\Delta=f_0$ or scan $\Delta$ with a fixed nonzero $\delta-\Delta=f_0$, the maximum transparency still occurs at $\delta=0$ for $\gamma_2=0$ due to the presence of a completely dark state. A nonzero $\gamma_2$ gradually shifts the transparency condition away from $\delta=0$. For scanning $\delta$ with $\gamma_2=0$, the maximum loss occurs at $g^2/f_0$, corresponding to one of the diagonalized states of the coupled $\ket{m_1}$ and $\ket{m_2}$ system. This effect also appears when scanning $\Delta$, where the two loss maxima are shifted to $\frac{-(\delta-\Delta)\pm \sqrt{\left(\delta-\Delta \right)^2+4g^2}}{2}$ due to the coupling between the two molecular states. In addition, the width gradually narrows as the detuning increases.

These effects are manifested in Fig.~\ref{sup2}, Fig.~\ref{sup3}, and Fig.~\ref{sup4}.

\subsection{Effects of Nonzero $W_2^{\rm eff}$ and Resonance Interference}
Now we focus on another level scheme where $W_2^{\rm eff}$ is nonzero. As discussed in Section~\ref{three-channel}, a nonzero $W_2^{\rm eff}$ (with $\gamma_2=0$) shifts the dark state condition from $\delta=0$ to $\delta=\frac{\left(V_{12}^{\rm eff}+G_{12}^{\rm eff}\right)w_2^{\rm eff}}{w_1^{\rm eff}}$. Meanwhile, a nonzero $W_2^{\rm eff}$ also causes asymmetric Fano line shapes and resonance interference \cite{ResInt}. 

\begin{figure*}[hbtp]
    \centering
    \includegraphics[width=0.9\textwidth]{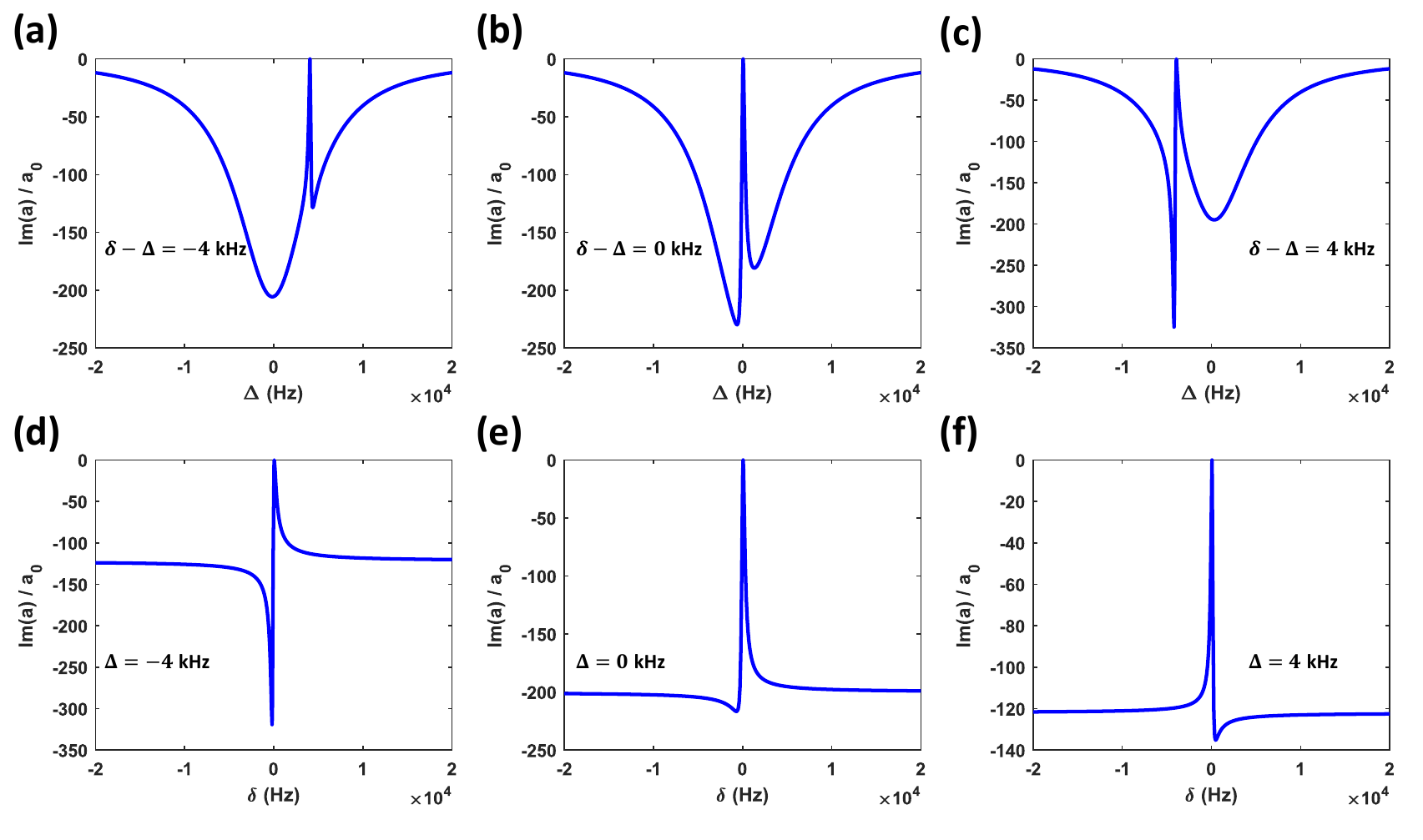}
    \caption{\textcolor{black}{Complex scattering length with a small $W_2^{\rm eff}$. Here, $E/h=10^{-12}$ Hz, $a_{bg}=160$ $a_0$, $g=-1$ kHz, $\gamma_1=10$ kHz, $\gamma_2=0$ kHz, $\Gamma_{10}^{\rm eff}=6.255$ kHz, and $\Gamma_{20}^{\rm eff}=0.0033\Gamma_{10}^{\rm eff}$. Panels (a)-(c) exhibit the complex scattering length as a function of $\Delta$ for fixed $\delta-\Delta=$ -4 kHz, 0 kHz, and 4 kHz, respectively. Panels (d)-(f) show the complex scattering length as a function of $\delta$ for fixed $\Delta=$ -4 kHz, 0 kHz, and 4 kHz, respectively. The dark state positions in (a)-(c) are shifted to $\left(-0.057-4.000\right)$ kHz, $\left(-0.057-0.000\right)$ kHz, and $\left(-0.057+4.000\right)$ kHz, respectively. Those in (d)-(f) are shifted to $\delta=-0.057$ kHz. The centers of the narrow Fano profiles in (a)-(c) are 4.100, 0, and -4.100 kHz, respectively, and those of the single Fano profiles in (d)-(f) are -0.098, 0, and 0.098 kHz, respectively. The additional coupling to the second molecular state makes the Fano line shapes in (b) and (e) asymmetric compared to the ideal cases in Fig.~\ref{sup3}(d) and Fig.~\ref{sup2}(d), respectively. Besides, the profiles in (a) and (c) are clearly not reflection-symmetric, and neither are those in (d) and (f). These asymmetries arise from resonance interference. The widths are 0.240 kHz, 0.417 kHz, and 0.240 kHz for (a)–(c), and 0.244 kHz, 0.400 kHz, and 0.244 kHz for (d)–(f). These values are the same as in the ideal case. The absolute local maxima in (a), (b) and (c) are $-0.306$ and $4.188$ kHz, $-1.316$ and $0.622$ kHz, and $-4.369$ and $0.163$ kHz, respectively. Those in (d), (e) and (f) are $-0.193$ kHz, $-0.696$ kHz, and $0.468$ kHz, respectively. All of them are shifted relative to the zero $W_2^{\rm eff}$ case.}}  
    \label{sup5}
\end{figure*}

As depicted in Fig.~\ref{sup5}, the dark state condition is shifted, and the Fano line shapes become asymmetric compared to those in Figs.~\ref{sup2} and~\ref{sup3}. Furthermore, the profiles in Fig.~\ref{sup5}(a) and Fig.~\ref{sup5}(c) are clearly not reflection-symmetric, especially near the absolute local maxima at $4.188$ kHz in (a) and $-4.369$ kHz in (c). The branch at $-4.369$ kHz in (a) is much enhanced, whereas the branch at $4.188$ kHz in (c) is suppressed. A similar effect is also manifested in Fig.~\ref{sup5}(d) and Fig.~\ref{sup5}(f). These phenomena are due to interference between the $\ket{m_1}$ induced resonance and the weak $\ket{m_2}$ induced resonance. In addition, the positions of the absolute local maxima are shifted relative to the zero $W_2^{\rm eff}$ case. However, the widths remain unchanged in the presence of $W_2^{\rm eff}$ when $\gamma_2=0$. The Fano profile centers remain symmetric under opposite detunings and, for most parameters, lie much closer to $\delta =0$ than the local loss minima.

These effects of nonzero $W_2^{\rm eff}$ are also observed in our experiment, as shown in Figs.~2 and~3 in the main text.

\subsection{Bound States in the Continuum (BICs)}
\begin{figure*}[t]
    \centering
    \includegraphics[width=1\textwidth]{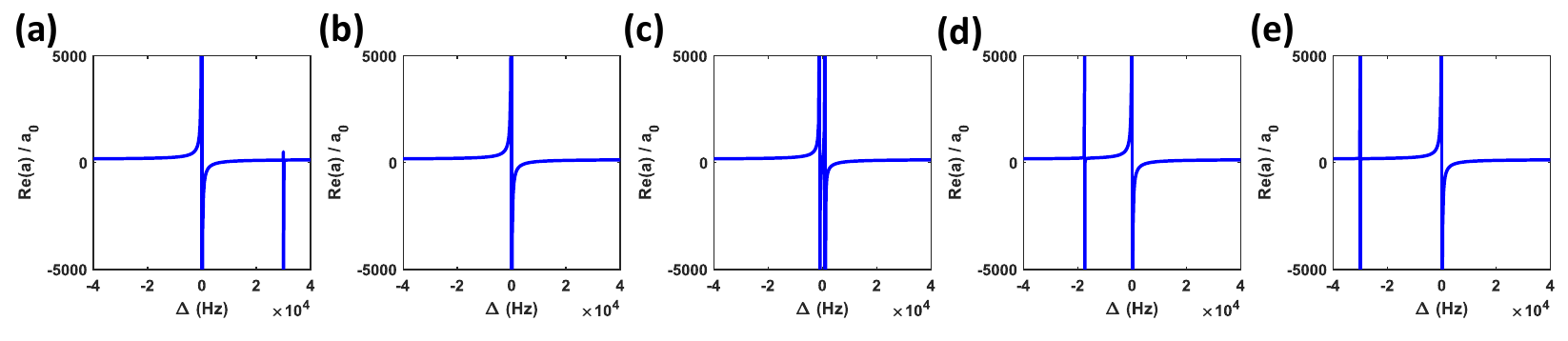}
    \caption{\textcolor{black}{Friedrich–Wintgen BICs from interference of two Feshbach resonances. Here, $E/h=10^{-12}$ Hz, $a_{bg}=160$ $a_0$, $g=-1$ kHz, $\gamma_1=0$ kHz, $\gamma_2=0$ kHz, $\Gamma_{10}^{\rm eff}=6.255$ kHz, $\Gamma_{20}^{\rm eff}=0.0033\Gamma_{10}^{\rm eff}$, $w_1^{\rm eff}=\sqrt{\Gamma_{10}^{\rm eff}}$, and $w_2^{\rm eff}=\sqrt{\Gamma_{20}^{\rm eff}}$. The imaginary part of the scattering length is zero and is therefore not shown. Panels (a)-(e) exhibit the real part of the scattering length as a function of $\Delta$ for fixed $\delta-\Delta$ values of -30 kHz, -17.350 kHz, 0 kHz, 17.350 kHz, and 30 kHz, respectively. A BIC occurs at $\delta-\Delta=g\frac{\left(w_1^{\rm eff}\right)^2 - \left(w_2^{\rm eff}\right)^2}{w_1^{\rm eff}w_2^{\rm eff}}=-17.350$ kHz, as shown in panel (b). At other detunings, there are two distinct resonance peaks in each panel.}}  
    \label{sup6}
    \vfill
    \centering
    \includegraphics[width=0.9\textwidth]{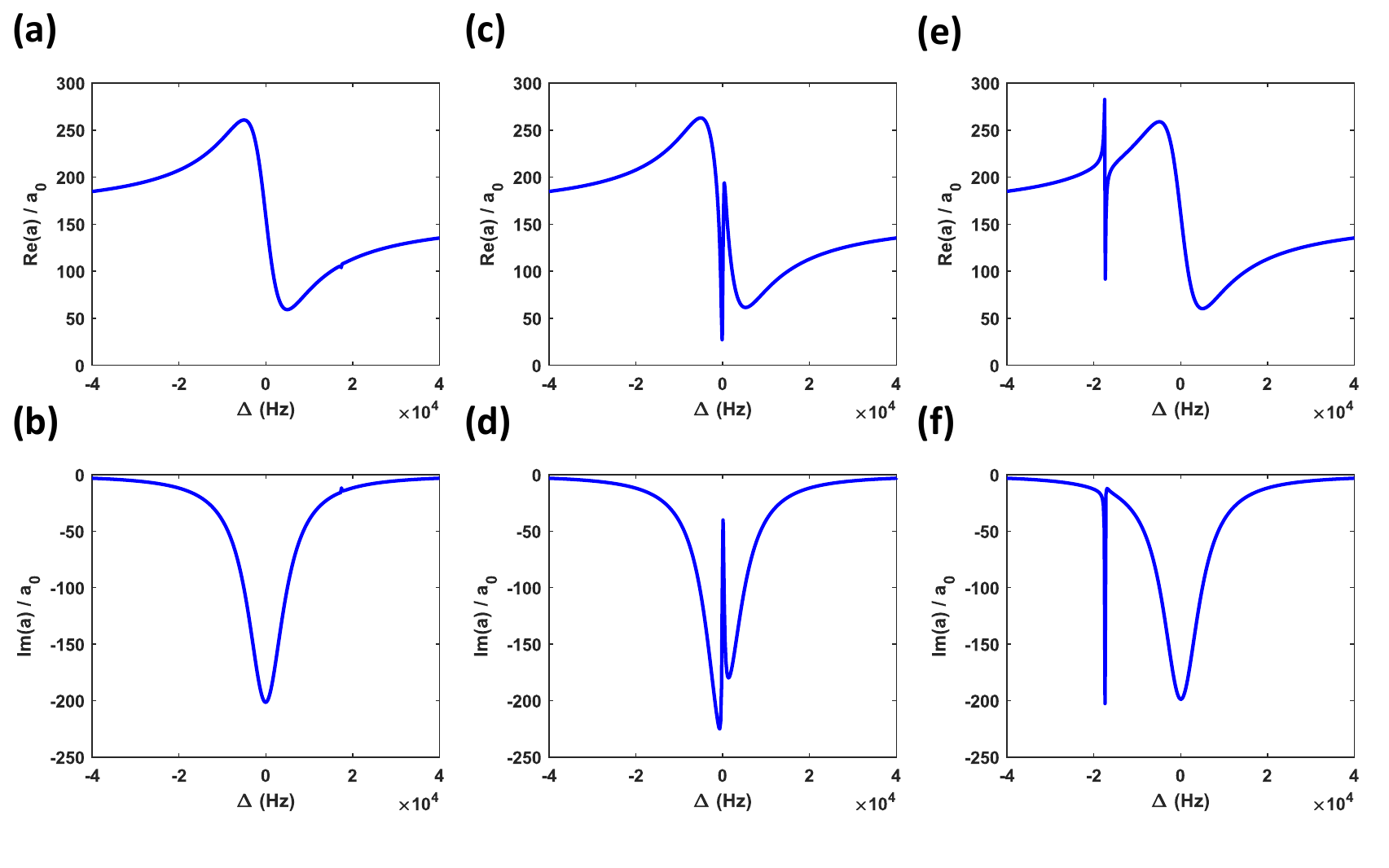}
    \caption{\textcolor{black}{BIC phenomena with nonzero decay rate. Here, $E/h=10^{-12}$ Hz, $a_{bg}=160$ $a_0$, $g=-1$ kHz, $\gamma_1=10$ kHz, $\gamma_2=0.1$ kHz, $\Gamma_{10}^{\rm eff}=6.255$ kHz, $\Gamma_{20}^{\rm eff}=0.0033\Gamma_{10}^{\rm eff}$, $w_1^{\rm eff}=\sqrt{\Gamma_{10}^{\rm eff}}$, and $w_2^{\rm eff}=\sqrt{\Gamma_{20}^{\rm eff}}$. Panels (a) and (b) depict the real and imaginary parts of the scattering length as a function of $\Delta$ at fixed $\delta-\Delta=-17.350$ kHz. Panels (c) and (d), and (e) and (f) show the complex scattering length at fixed $\delta-\Delta=0$ kHz and $17.350$ kHz, respectively. At the BIC condition, one of the Fano profiles is almost completely suppressed in both the real and imaginary parts of the scattering length, as shown in (a) and (b).}}  
    \label{sup7}
\end{figure*}

Interfering two resonances can also lead to bound states in the continuum (BICs) \cite{ResInt,BICs}. The condition for such BICs (with $\gamma_1=0$ and $\gamma_2=0$) is \cite{ResInt}
\begin{equation}
    \Delta-\delta= \left(V_{12}^{\rm eff}+G_{12}^{\rm eff}\right)\frac{\left(w_1^{\rm eff}\right)^2 - \left(w_2^{\rm eff}\right)^2}{w_1^{\rm eff}w_2^{\rm eff}},
\end{equation}
which depends on the relative signs of $w_1^{\rm eff}$, $w_2^{\rm eff}$, and $V_{12}^{\rm eff}+G_{12}^{\rm eff}$. Figure~\ref{sup6} shows an example of this kind of BIC. In panels (a) and (c)-(e), there are two Feshbach resonance peaks, while in panel (b), the BIC condition is satisfied and one of the resonances disappears, indicating the formation of a BIC.

When $\gamma_1$ and $\gamma_2$ are nonzero, this effect still persists near the BIC condition given above and is manifest in both the real and imaginary parts of the scattering length. As depicted in Fig.~\ref{sup7}(a) and (b), one of the Fano profiles is almost completely suppressed in both the real and imaginary parts of the scattering length.

The BIC effect can also be observed in the complex scattering length as a function of $\delta$ with fixed $\Delta$. Specifically, the presence of a BIC makes the scattering length nearly unchanged when scanning $\delta$ at $\Delta=\left(V_{12}^{\rm eff}+G_{12}^{\rm eff}\right)\frac{w_1^{\rm eff}}{w_2^{\rm eff}}$.

\subsection{Experimental Observation of BIC Effects}
\begin{figure*}[h]
    \centering
    \includegraphics[width=1.0\textwidth]{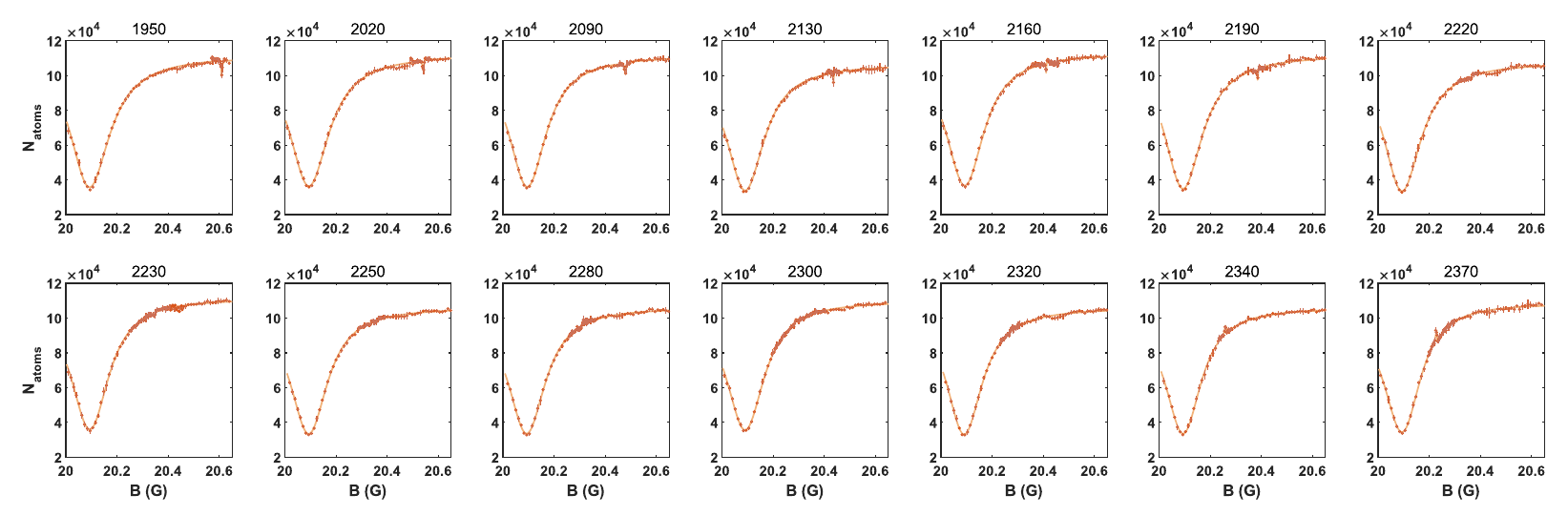}
    \caption{\textcolor{black}{Experimental BIC signals from magnetic field scan. The atom loss measurement over a wider range of modulation frequencies (1.950–2.370 MHz) is implemented as a complement to Fig.~3 (2.400–2.600 MHz) in the main text. The title of each panel denotes the modulation frequency in units of kHz. The error bars are standard errors, and the data are fitted with a sum of Fano line shapes. Two distinct Fano profiles are observed at modulation frequencies of 1950, 2020, 2090, 2130, 2160, 2190, 2340, and 2370 kHz, while the narrow branch is strongly suppressed at 2220, 2230, 2250, 2280, and 2320 kHz.}}  
    \label{sup8}
    \vfill
    \centering
    \includegraphics[width=0.7\textwidth]{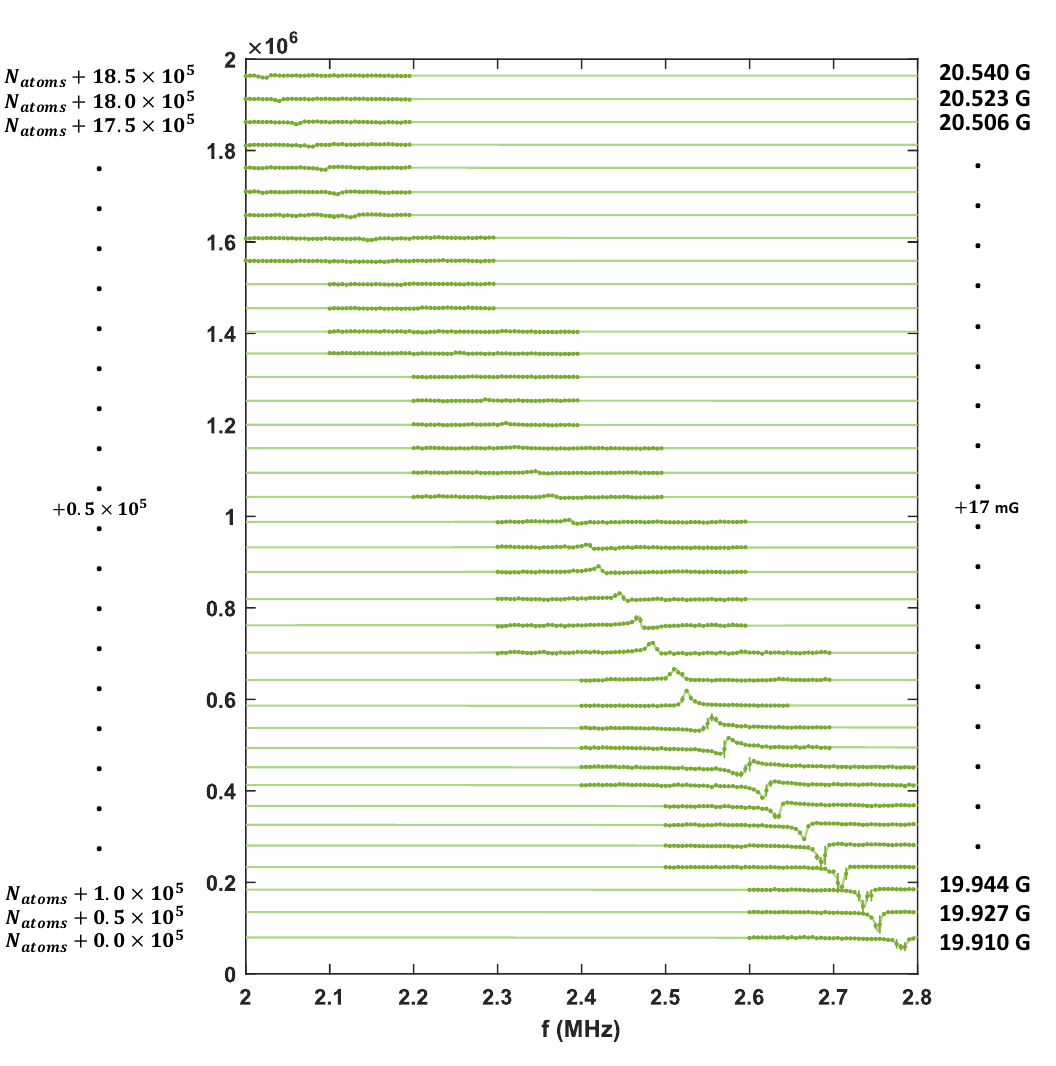}
    \caption{\textcolor{black}{Experimental BIC signals from modulation frequency scan. The error bars represent standard errors, and the data are fitted with a single Fano line shape. Each loss profile is obtained at a fixed magnetic field, ranging from 19.910 G to 20.540 G in steps of 0.017 G. These profiles are plotted in a common panel, with an artificial atom number spacing of $0.5\times10^5$ between adjacent magnetic field values. In the magnetic field range of 20.250–20.404 G, the atom number remains nearly unchanged when scanning the modulation frequency, while for other field values, a distinct Fano profile is observed.}}  
    \label{sup9}
\end{figure*}

Here, we extend the loss measurement Fig.~3 (2.400-2.600 MHz) in the main text to a wider range of modulation frequencies (1.950-2.370 MHz). As shown in Fig.~\ref{sup8}, one of the Fano branches of the measured atom loss profile is greatly suppressed in the modulation frequency range of 2.220-2.320 MHz. This signal corresponds to a BIC, as discussed in the last subsection.

Figure~\ref{sup9} shows the measured atom loss signals as a function of the modulation frequency at different values of the fixed magnetic field, corresponding to the scanning $\delta$ with fixed $\Delta$. In the range of 20.250 to 20.404 G, the atom number remains nearly unchanged while scanning the modulation frequency. This is also a signature of a BIC.

Using the experimentally measured parameters, the next section will show that these ranges are indeed near the BIC conditions.

\section{Determination of Experimental Parameters}
\label{section4}
In this section, we describe how the key parameters entering the effective Hamiltonian and the scattering model are determined experimentally. These parameters include the modulation amplitudes, the magnetic moment of the molecular state, the effective coupling strengths, and the condition for BIC formation. Establishing these quantities is essential for connecting the theoretical description developed above to the measured atom-loss spectra. In particular, the calibration of the modulation amplitudes allows us to determine the relevant Bessel-function renormalization factors, while the measured magnetic moment converts the resonance-frequency dependence into magnetic-field detuning. 

\subsection{Experimental Implementation of the Modulation Protocol}
As mentioned in the main text, we use the AC Stark effect of a laser to shift the energy between the two molecular states, as well as between them and the free-atom state. Thus, by modulating the light intensity, we achieve the desired modulation of these states. 

Light intensity modulation is achieved through the interference of two laser beams, each passing through a separate acousto-optic modulator (AOM) operating at different frequencies, combined by a beam splitter. For the single-frequency modulation in the main text, the modulation frequency is controlled by fixing one AOM at +83 MHz while tuning the driving frequency of the other AOM. To introduce a second modulation frequency, we add another pair of AOMs. The second modulation frequency is controlled by fixing one AOM at -75 MHz while tuning the driving frequency of the other AOM. The variation of light intensity over time is measured by a biased Silicon detector with a bandwidth of about 80~MHz, which shows that the half-width of the modulation frequency is lower than 1 Hz. In addition to this AC component, modulating the optical intensity in this way also introduces an inevitable DC component of the light shift. The diameter of the first beam spot at the atomic position is about 700~$\mu$m and the diameter of the second beam spot at the atomic position is about 650~$\mu$m, so it is guaranteed that the light intensity at the atomic position is uniform.

For this implementation, the modulation amplitudes $\alpha_i$ and $\beta_i$ ($i$= 0, 1, 2) in Eq.~\ref{amplitude} increase linearly with optical intensity with a slope that depends on the laser frequency, allowing them to be tuned by varying both parameters. In the single-frequency modulation experiment (including Figs.~2, 3, and 4 in the main text), the laser frequency (before passing through the AOM) is set to 351.45420 THz. At this frequency, the differential light shift between the free-atom state and the $4g4$ molecular state (i.e. $\ket{m_1}$) is large, while that between the free-atom state and the $6s$ molecular state (i.e. $\ket{m_2}$) is kept small. For the dual-frequency modulation experiment (including Fig.~5 in the main text), the laser frequency (before passing through the AOM) is set to 351.58428 THz. At this frequency, the differential light shift between the free-atom state and the $6s$ state has the same order of magnitude as that for the $4g4$ state, but with the opposite sign, allowing the construction of various level schemes. 

In the range of a few hundred GHz detuning relative to the cesium D2 transition used in this work, not only atomic transitions, but also molecular state transitions contribute to the AC Stark shift~\cite{Ref1,Ref5,shift1,shift2}. In spite of this, we can directly measure the light shift in the experiment, as discussed in the next subsection. In addition, inelastic processes through bound-bound transitions may occur in this detuning range~\cite{Ref5}. As mentioned in Section~\ref{three-channel}, we treat this process, along with the inelastic process of scattering into higher Floquet modes and other inelastic mechanisms, as a phenomenological decay rate of the corresponding molecular state. These processes can be strengthened by increasing the modulation intensity, leading to an increase in the corresponding decay rates. The Fano widths of the measured loss profiles qualitatively reflect the decay rates $\gamma_1$ and $\gamma_2$.

\subsection{Measurement of Modulation Amplitudes}
We calibrate the modulation amplitudes in our experiment using three methods. First, for a molecular state that exhibits a magnetic Feshbach resonance, the differential light shift between this state and the free-atom state can be determined by comparing the shifted resonance magnetic field position with the original resonance position. Then, we can quantify the differential modulation amplitudes using the measured intensity modulation depth. For example, Fig.~2(a) in the main text shows a shifted magnetic Feshbach resonance of $4g4$ centered near 20.09 G. Compared to the original resonance position of 19.84 G, the DC component of the applied intensity-modulated laser shifts it by 0.25 G, which corresponds to 199 kHz with $\Delta\mu_1=0.797(1)$ MHz/G. Since the modulation depth is nearly 1, the differential modulation amplitude between the $4g4$ state and the free-atom state is 199(8) kHz.

\begin{figure*}[h]
    \centering
    \includegraphics[width=0.8\textwidth]{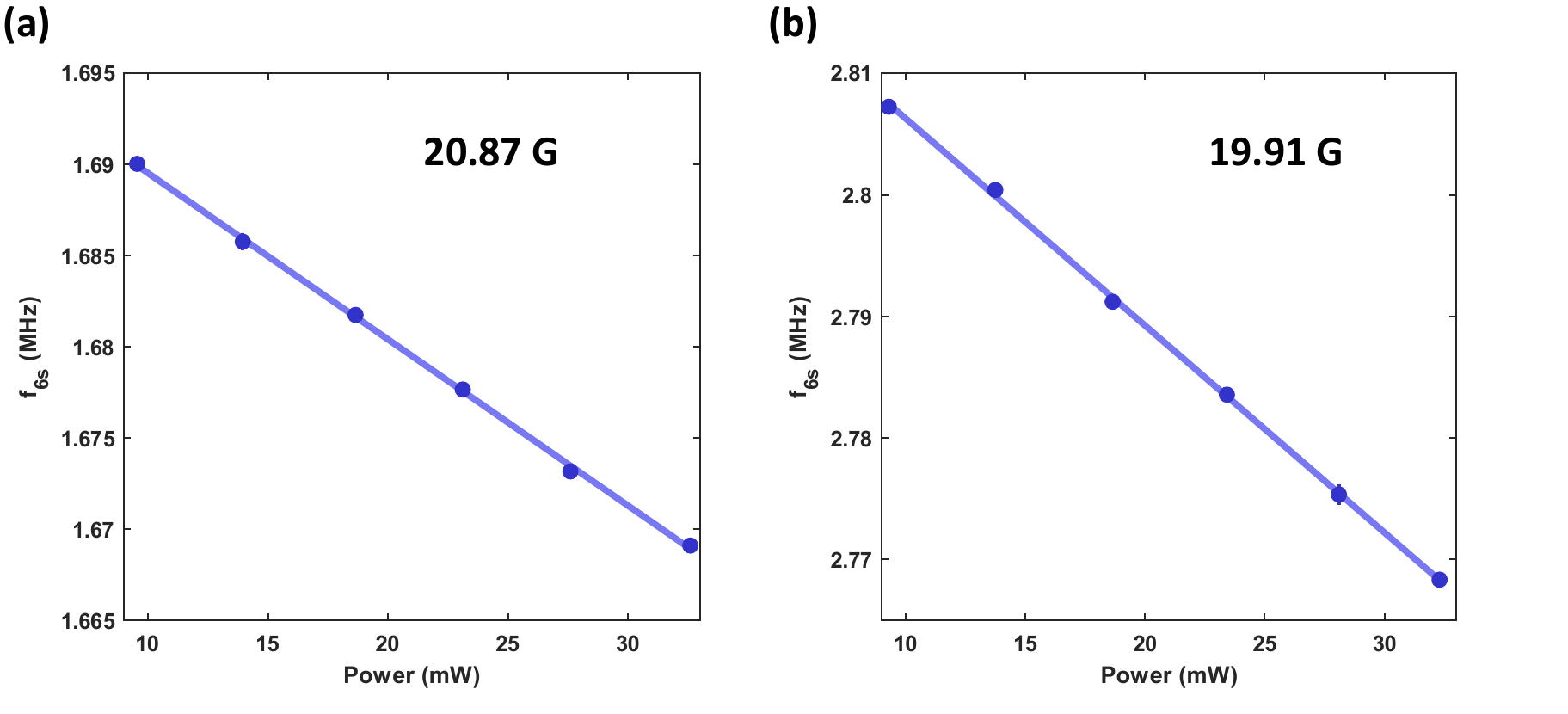}
    \caption{\textcolor{black}{The measured resonance frequency positions of $6s$ state as a function of the power of the intensity-modulated laser. The data in (a) and (b) are linearly fitted. (a) At 20.87 G, the fitted slope is -0.91(1) kHz/mW and the intercept is 1698.6(3) kHz. (b) At 19.91 G, the fitted slope is -1.70(2) kHz/mW and the intercept is 2823.3(4) kHz.}} 
    \label{sup10}
\end{figure*}

Second, for a general molecular state that may be far from the magnetic resonance position within the applicable magnetic field range, we can determine the modulation amplitudes by fitting the measured resonance positions of first-order modulation-induced Feshbach resonances, following the method used in Ref.~\cite{Ref1}. Fig.~\ref{sup10} shows an example for the $6s$ state, where the measured resonance frequency positions are fitted linearly as a function of the power of the intensity-modulated laser. The fitted slopes are -0.91(1) kHz/mW and -1.70(2) kHz/mW, and the intercepts are 1698.6(3) kHz and 2823.3(4) kHz for panels (a) and (b) at 20.87 G and 19.91 G, respectively. Thus, for the 18.60(5) mW power used in Figs.~2 and 3 in the main text, the differential modulation amplitudes between the $6s$ state and the free-atom state are determined by multiplying the fitted slopes by the power, yielding -16.9(2) kHz at 20.87 G and -31.6(4) kHz at 19.91 G, respectively. The same calibration procedure can be applied to the $4g4$ state.

Third, we can use modulation interferograms to extract the amplitudes by simply scanning the modulation frequency, without needing to vary the optical intensity. This will be reported in a future work.

In summary, the measured modulation amplitudes for the setup in the main text are listed as follows.

The differential modulation amplitude per mW between the $4g4$ state and the free-atom state in Fig.~2 and 3 of the main text is 10.7(4) kHz/mW. The differential modulation amplitude per mW between the $6s$ state and the free-atom state in Fig.~2 and 3 is -1.59(5) kHz/mW. Thus, The differential modulation amplitude per mW between the $4g4$ state and the $6s$ state in Fig.~2 and 3 is 12.3(4) kHz/mW. The modulation beam has a diameter of approximately 700~$\mu$m at the atomic position, with a total power of 18.6 mW.

In Fig.~4 of the main text, the differential modulation amplitude per mW between the $4g4$ state and the free-atom state is 10.7(4) kHz/mW. The differential modulation amplitude per mW between the $6s$ state and the free-atom state lies in the range of (-1.7,-1.2) kHz/mW over the magnetic field range of 19.9-20.6 G. The modulation beam has a diameter of approximately 700~$\mu$m at the atomic position, with a total power of 9.47, 12.80, 16.15, 18.60, 21.50, 26.60, 33.60 and 39.80 mW, respectively.

In Fig.~5 of the main text, the differential modulation amplitude per mW between the $4g4$ state and the free-atom state is 23.3(7) kHz/mW for the first modulation frequency $\omega_{\rm I}$ and 10.2(4) kHz/mW for the second modulation frequency $\omega_{\rm II}$. The differential modulation amplitude per mW between the $6s$ state and the free-atom state is -4.19(5) kHz/mW, -3.7(1) kHz/mW, and -3.2(1) kHz/mW for the first modulation frequency $\omega_{\rm I}$ at 20.35 G, 20.44 G, and 20.57 G, respectively, and -5.58(5) kHz/mW, -5.3(1) kHz/mW, and -5.0(1) kHz/mW for the second modulation frequency $\omega_{\rm II}$ at the same magnetic field values, respectively. The diameter of the first beam spot at the atomic position is about 700~$\mu$m and the diameter of the second beam spot at the atomic position is about 650~$\mu$m. The power is 9.80(5) mW for the first tone and 10.00(5) mW for the second tone.

\subsection{Measurement of Magnetic Moment}
We determine the differential magnetic moment $\Delta\mu_i$ between the molecular state $\ket{m_i}$ and the free-atom state (or equivalently, the energy spectrum) using the modulation spectroscopy as in Ref.~\cite{Ref1}. The procedure is essentially the same as the second method described in the previous subsection, wherein the fitting intercepts are used as the unshifted binding energy.

The measured $\Delta\mu_1$ is 0.797(1) MHz/G. $\Delta\mu_2$, while generally field-dependent, is approximately constant for the range of 20.05–20.20 G in Fig.~3 of the main text, with a measured value of 1.238(8) MHz/G. The measured energy spectrum of $\ket{m_2}$ is plotted as a black dot-dashed line in Fig.~3(c) of the main text. 

\subsection{Quantifying the Coupling Terms}
We quantify the coupling terms of $\Gamma_{10}^{\rm eff}$, $\Gamma_{20}^{\rm eff}$, and $V_{12}^{\rm eff}+G_{12}^{\rm eff}$ (as defined in section~\ref{three-channel}) by performing coupled-channel calculations. More explicitly, we first theoretically determine the original coupling terms (i.e., $\Gamma_{10}$, $\Gamma_{20}$, and $V_{12}+G_{12}$ without modulation) from coupled-channel calculations, and then use the measured modulation amplitudes to quantify the experimental effective coupling terms based on the corresponding experimental modulation factors.

\begin{figure*}[hbtp]
    \centering
    \includegraphics[width=1\textwidth]{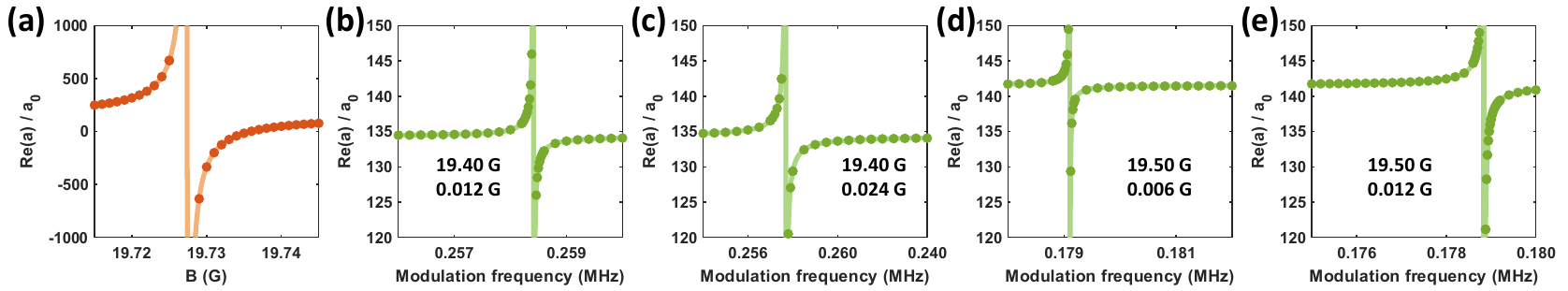}
    \caption{\textcolor{black}{Real part of the scattering length for the cesium $4g4$ Feshbach resonance from coupled-channel calculation. (a) Magnetic Feshbach resonance near 19.7275 G. The resonance position deviates slightly from the experimental value of 19.84 G due to the limited precision of the potential parameters. Fitting the calculated data to a Breit-Wigner formula yields $a_{bg}=150 a_0$ and $\Delta_{B}=8.5$ mG, from which we obtain $a_{bg}\Gamma_{10}/h = 150a_0 \times 7$ kHz (the calculated $\Delta\mu_1$ under the corresponding potential parameters is $0.8$ MHz/G). A more accurate calculation would require further refinement of the potential parameters. (b)-(e) The modulation-induced $4g4$ Feshbach resonance with an oscillating magnetic field $B_{\rm rf}=0.012,~0.024,~0.006,~0.012$ G at a bias field of 19.40, 19.40, 19.50 and 19.50 G, respectively. The data are also obtained from coupled-channel calculations and fitted with a Breit–Wigner form. The fitted values of $a_{bg}\Gamma_{10}^{\rm eff}/h$ are $134a_0\times2.9$ Hz, $134a_0\times11.5$ Hz, $142a_0\times1.36$ Hz and $142a_0\times5.40
    $ Hz. They are close to the corresponding values of $134a_0\times2.7$ Hz, $134a_0\times10.8$ Hz, $142a_0\times1.32$ Hz and $142a_0\times5.29$ Hz by multiplying $a_{bg}\Gamma_{10}/h$ with modulation factor $\left|J_1(\frac{\alpha_1-\alpha_0}{\omega}) \right|^2$. }} 
    \label{sup13}
\end{figure*}

For $\Gamma_{10}$, we fit the calculated real part of the scattering length as a function of the magnetic field near the $4g4$ Feshbach resonance using a Breit–Wigner form
\begin{equation}
    \mathrm{Re}(a) = a_{bg} \left(1-\frac{\left(\epsilon-\epsilon_0\right)\Gamma_{0}}{\left(\epsilon-\epsilon_0 \right)^2+\left(\gamma/2 \right)^2} \right),
\end{equation}
where $\Gamma_0$ denotes $\Gamma_{i0}$ or $\Gamma^{\rm eff}_{i0}$, $\gamma$ is the decay rate, and $\left(\epsilon-\epsilon_0\right)$ represents $\Delta\mu(B-B_0)$ for the magnetic Feshbach resonance. Using the cesium interaction potentials in Ref.~\cite{2012model}, the extracted $a_{bg}\Gamma_{10}/h$ for the $4g4$ Feshbach resonance is $150a_0 \times 7$ kHz, as shown in Fig.~\ref{sup13}(a). 

For $\Gamma_{20}$, we first calculate the scattering length using a modified Hamiltonian that includes an additional oscillating magnetic field term. This modulation term can induce a Feshbach resonance of the $6s$ state when the modulation frequency of the oscillating magnetic field matches the binding energy of the $6s$ state. We then extract $\Gamma_{20}^{\rm eff}$ for this induced $6s$ Feshbach resonance using the Breit–Wigner fit. The original $\Gamma_{20}$ is then determined by dividing the extracted $\Gamma_{20}^{\rm eff}$ by the modulation factor of the oscillating magnetic field.

Specifically, the Hamiltonian with an additional oscillating magnetic field term, along with its matrix elements in a decoupled basis, can be found in Refs.~\cite{Ref1,2012model,Mag,Mag2}. The scattering calculation uses the same propagation methods as in Ref.~\cite{2012model}. The molecular angular momentum of their relative motion, $L$, is truncated to 4, and the photon number (or drive quantum number) $N$ is truncated to 1 for the small oscillating magnetic field values used below. 

\begin{figure*}[hbtp]
    \centering
    \includegraphics[width=1\textwidth]{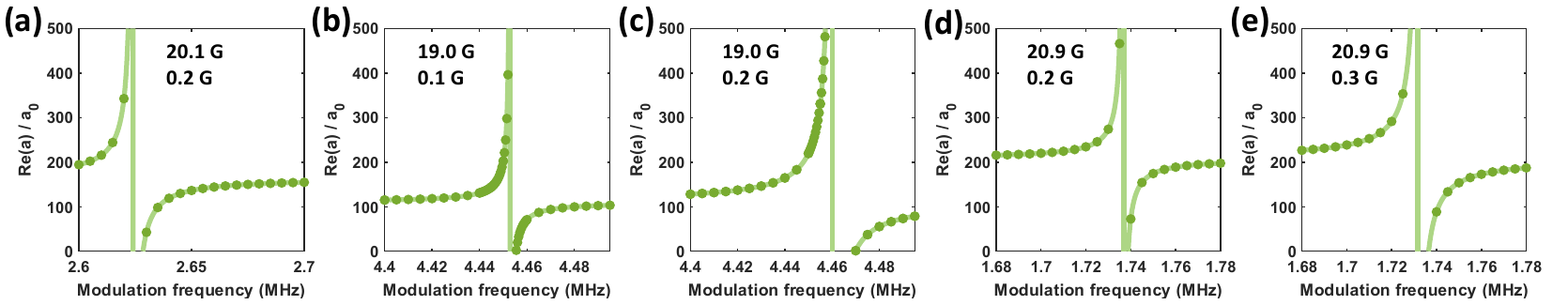}
    \caption{\textcolor{black}{Calculated real part of the scattering length as a function of modulation frequency for the cesium $6s$ Feshbach resonance induced by an oscillating magnetic field. Panels (a)–(e) show five modulation-induced Feshbach resonances of the $6s$ state at bias magnetic fields of 20.1 G (a), 19.0 G (b), 19.0 G (c), 20.9 G (d) and 20.9 G (e), with corresponding $B_{\rm rf}$ values of 0.2, 0.1, 0.2, 0.2 and 0.3 G, respectively. The data are obtained from coupled-channel calculations and fitted with a Breit–Wigner form. The fitted values of $a_{bg}\Gamma_{20}^{\rm eff}$ are $164a_0 \times 4.4$ kHz, $110a_0 \times 2.5$ kHz, $110a_0 \times 10.0$ kHz, $208a_0 \times 2.1$ kHz and $208a_0 \times 4.7$ kHz, respectively, which can be used to estimate $a_{bg}\Gamma_{20}$ by dividing the corresponding modulation factor. The fitted $a_{bg}\Gamma_{20}^{\rm eff} \propto B_{\rm rf}^2$ (e.g., (b)/(c) and (d)/(e)), consistent with $\left|J_1(\frac{\alpha_2-\alpha_0}{\omega}) \right|^2$ at small $\frac{\alpha_2-\alpha_0}{\omega}$.}} 
    \label{sup11}
\end{figure*}

Figure~\ref{sup11} shows modulation-induced Feshbach resonances of the cesium $6s$ state with an oscillating magnetic field $B_{\rm rf}$ at bias magnetic fields of 20.1 G (a), 19.0 G (b), 19.0 G (c), 20.9 G (d), and 20.9 G (e), with corresponding $B_{\rm rf}$ values of 0.2, 0.1, 0.2, 0.2, and 0.3 G. The calculated data are fitted with a Breit–Wigner form, where the fitted $a_{bg}\Gamma_{20}^{\rm eff}$ can be used to estimate $a_{bg}\Gamma_{20}$ by dividing the modulation factor $\left|J_1(\frac{\alpha_2-\alpha_0}{\omega}) \right|^2$. Here, $\omega$ is the modulation angular frequency of the oscillating field, and $\alpha_2-\alpha_0$ is estimated as $\Delta\mu_2B_{\rm rf}/\hbar$ (this approximation requires a relatively small $B_{\rm rf}$). For a small $B_{\rm rf}$, we find that the fitted $a_{bg}\Gamma_{20}^{\rm eff} \propto B_{\rm rf}^2$ when varying $B_{\rm rf}$ (such as in Fig.~\ref{sup11}(b) and (c), Fig.~\ref{sup11}(d) and (e)), consistent with $\left|J_1(\frac{\alpha_2-\alpha_0}{\omega}) \right|^2$ at small $\frac{\alpha_2-\alpha_0}{\omega}$. To estimate the actual experimental $a_{bg}\Gamma_{20}^{\rm eff}$, we multiply the obtained $a_{bg}\Gamma_{20}$ by the modulation factor derived from the measured modulation amplitude.

The above procedure also allows us to verify that the effective coupling term after modulation is well approximated by the original term multiplied by the corresponding Bessel modulation factor. Specifically, we compute the scattering length for the induced $4g4$ Feshbach resonance, extract $\Gamma_{10}^{\rm eff}$ and compare it to the previously fitted value of $\Gamma_{10}$. As shown in Fig.~\ref{sup13}(b)-(e), the extracted values of $a_{bg}\Gamma_{10}^{\rm eff}/h$ are $134a_0\times2.9$ Hz, $134a_0\times11.5$ Hz, $142a_0\times1.36$ Hz, and $142a_0\times5.40$ Hz, respectively. They are close to the corresponding values of $134a_0\times2.7$ Hz, $134a_0\times10.8$ Hz, $142a_0\times1.32$ Hz, and $142a_0\times5.29$ Hz by multiplying $a_{bg}\Gamma_{10}/h$ with the modulation factor $\left|J_1(\frac{\alpha_1-\alpha_0}{\omega}) \right|^2$. In addition, we note that when the magnetic field is tuned too far from the magnetic Feshbach resonance of $4g4$, while the calculated scattering length remains well captured by the Breit-Wigner formula, the extracted $a_{bg}\Gamma_{10}^{\rm eff}$ gradually deviates from the prediction obtained by multiplying the fitted $a_{bg}\Gamma_{10}$ (obtained near 19.7275 G) by the Bessel modulation factor. Meanwhile, the scaling $a_{bg}\Gamma_{10}^{\rm eff} \propto B_{\rm rf}^2$ still holds. This behavior is attributed to multichannel effects, which may alter the original term $a_{bg}\Gamma_{10}$ at these magnetic field values while preserving the scale of the Bessel modulation factor. These effects are even more pronounced when extracting $a_{bg}\Gamma_{20}$, which generally varied with the magnetic field. Therefore, to estimate the experimental values of $a_{bg}\Gamma_{20}^{\rm eff}$, we should use $a_{bg}\Gamma_{20}$ extracted at the corresponding experimental magnetic field.

\begin{figure*}[hbtp]
    \centering
    \includegraphics[width=0.95\textwidth]{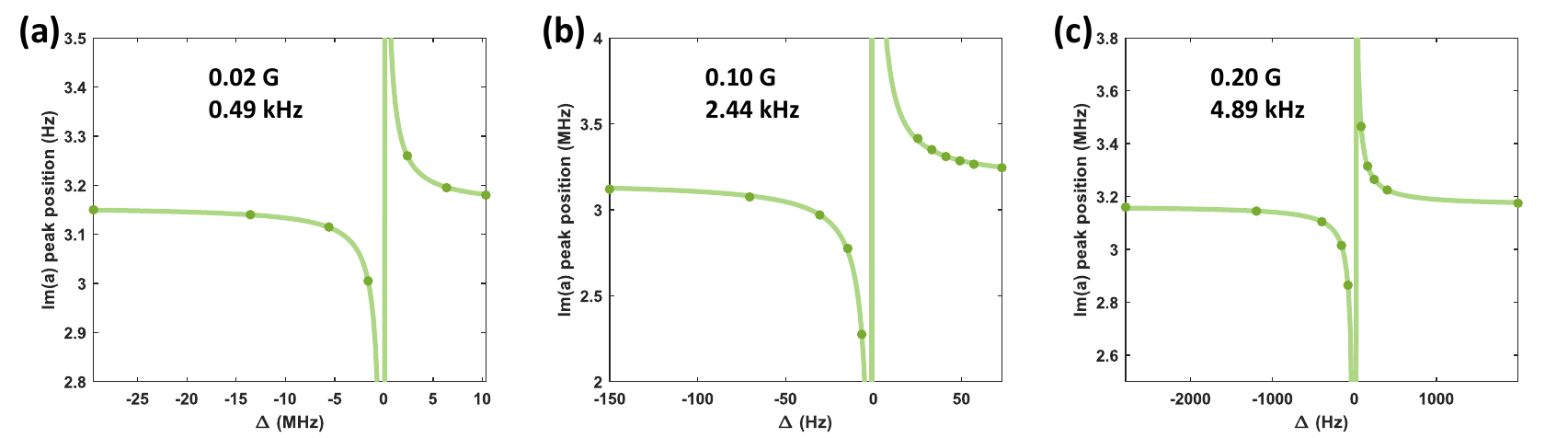}
    \caption{\textcolor{black}{The extracted peak positions of $\left|\mathrm{Im(a)}\right|$ as a function of $\Delta$. The data are extracted from the coupled-channel calculation and fitted to $\frac{\left( V_{12}^{\rm eff}+G_{12}^{\rm eff} \right)^2}{\Delta}$. (a) $B_{\rm rf}=0.02$ G and the fitted $\left| V_{12}^{\rm eff}+G_{12}^{\rm eff} \right|$ is 0.49 kHz. (b) $B_{\rm rf}=0.10$ G and the fitted $\left| V_{12}^{\rm eff}+G_{12}^{\rm eff} \right|$ is 2.44 kHz. (c) $B_{\rm rf}=0.20$ G and the fitted $\left| V_{12}^{\rm eff}+G_{12}^{\rm eff} \right|$ is 4.89 kHz. The fitted values of $\left| V_{12}^{\rm eff}+G_{12}^{\rm eff} \right|$ scale linearly with $B_{\rm rf}$, as expected from $\left| V_{12}^{\rm eff}+G_{12}^{\rm eff} \right| = \left|J_1(\frac{\alpha_2-\alpha_1}{\omega}) \right| \left| V_{12}+G_{12} \right|$ for small $\frac{\alpha_2-\alpha_1}{\omega}$.}} 
    \label{sup12}
\end{figure*}

For $V_{12}+G_{12}$, we also utilize the coupled-channel calculation with an oscillating magnetic field. Here, we calculate the scattering length as a function of modulation frequency near the $4g4$ Feshbach resonance, where the bias magnetic field is set close to the theoretical value of 19.7275 G (slightly deviating from the experimental value of 19.84 G, depending on the potential parameters). When the modulation frequency is near the binding energy of the $6s$ state, a three-level configuration forms. For the modulation parameters used below, the decay rates $\gamma_1$ and $\gamma_2$ are small, and the absolute maxima of $\mathrm{Im}(a)$ satisfy $\frac{\left( V_{12}^{\rm eff}+G_{12}^{\rm eff} \right)^2}{\Delta}$ when we scan the modulation frequency at a fixed magnetic field with a $4g4$ detuning $\Delta$, as discussed in Section~\ref{section3} and Fig.~\ref{sup2}. Therefore, we vary the magnetic field near 19.7275 G to tune the $\Delta$ and extract, for each $\Delta$, the modulation frequencies at which the absolute maximum of $\mathrm{Im}(a)$ occurs. Subsequently, we can determine $\left| V_{12}^{\rm eff}+G_{12}^{\rm eff} \right|$ by fitting the extracted frequencies as a function of $\Delta$ to $\frac{\left( V_{12}^{\rm eff}+G_{12}^{\rm eff} \right)^2}{\Delta}$.

Figure~\ref{sup12} shows the extracted positions as a function of $\Delta$ for $B_{\rm rf}=$ 0.02, 0.10, and 0.20 G, with the corresponding fitted values of $\left| V_{12}^{\rm eff}+G_{12}^{\rm eff} \right|$ being 0.49, 2.44, 4.89 kHz, respectively. These values scale linearly with $B_{\rm rf}$, as expected from $\left| V_{12}^{\rm eff}+G_{12}^{\rm eff} \right| = \left|J_1(\frac{\alpha_2-\alpha_1}{\omega}) \right| \left| V_{12}+G_{12} \right|$ for small $\frac{\alpha_2-\alpha_1}{\omega}$. Here, $\alpha_2-\alpha_1$ is estimated as $\left(\Delta\mu_2-\Delta\mu_1 \right)B_{\rm rf}/\hbar$ for small $B_{\rm rf}$. Using the given $\Delta\mu_1$ and $\Delta\mu_2$,   $\left| V_{12}+G_{12} \right|$ is estimated to be approximately 200 kHz.

\begin{figure*}[hbtp]
    \centering
    \includegraphics[width=0.4\textwidth]{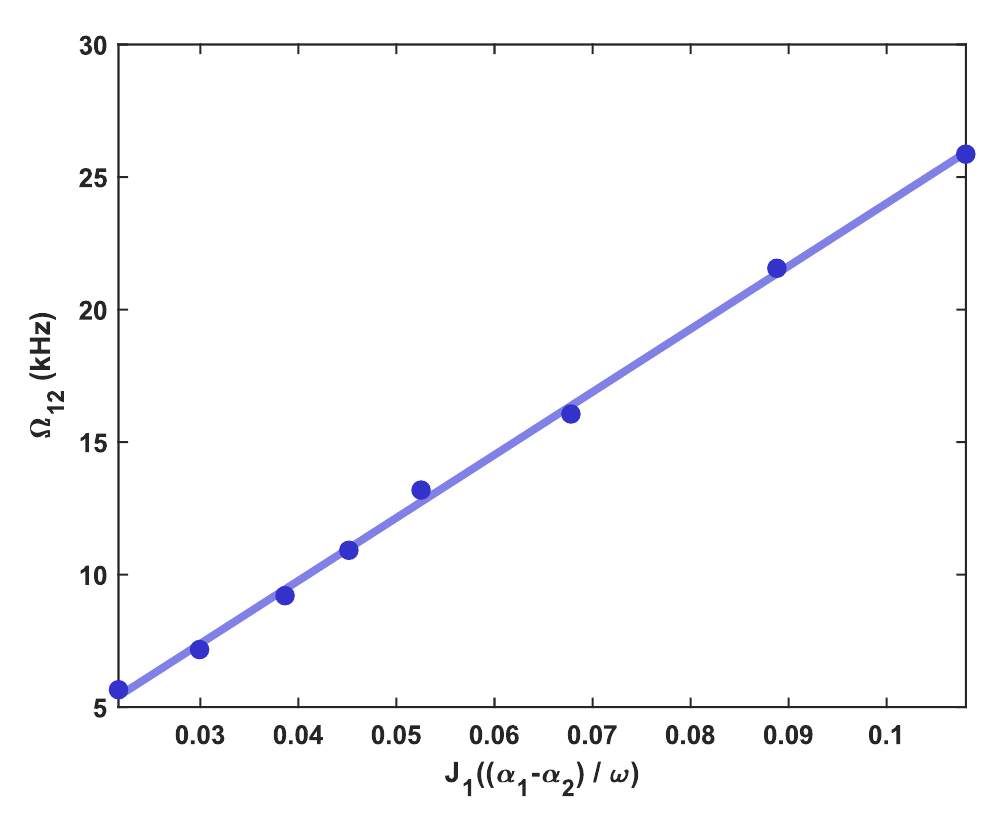}
    \caption{\textcolor{black}{Experimental data for the extracted half-distance of local loss maxima from Fig.~4 of the main text (in frequency units) versus the measured modulation amplitude expressed as a modulation factor. $\Omega_{12}$ equals the extracted half-distance times $\Delta\mu_1/h=0.797$ MHz/G.The data are fitted to a linear function, resulting in a slope of 240 kHz and an intercept of 0 kHz.}} 
    \label{sup14}
\end{figure*}

It is important to note that $\left| V_{12}+G_{12} \right|$ may gradually deviate from the above estimated value as the bias magnetic field deviates from 19.7275 G, due to the curvature of the $6s$ state~\cite{2012model}. For the magnetic field range considered in the main text, the deviation should be relatively small. For comparison with our experiment, we extract the spacing between the two local minima in each loss profile presented in Fig.~4(a) of the main text. As discussed in Section~\ref{section3}, this distance corresponds to $2\left| V_{12}+G_{12} \right|$ in the ideal case. However, this distance can be broadened by a nonzero $\gamma_2$ and a relatively strong $\gamma_1$. A nonzero difference $\Delta\mu_1-\Delta\mu_2$ can broaden or narrow this distance depending on its sign, since $\delta$ also varies as we tune $\Delta$ by changing the magnetic field. We quantify these effects by varying the corresponding parameters in Eq.~\ref{eq36} (or equivalently, the MQDT parameters discussed in Section~\ref{section2}). Given the parameter range determined from coupled-channel calculations and experiment, the theoretical distances agree with $2\left| V_{12}+G_{12} \right|$ to within a relative error of $\pm30$\% when these effects are included.

Figure~\ref{sup14} displays the experimental data for the extracted half-distance (in frequency units) versus the measured modulation amplitude, expressed as a modulation factor, where $\Omega_{12}$ equals the extracted half-distance times $\Delta\mu_1/h=0.797$ MHz/G. The data are linearly fitted, yielding a slope value of $240$ kHz. As expected, this value is close to the estimated value of 200 kHz obtained from coupled-channel calculations. In addition, we repeat the same measurement at other laser detunings, where the ratio of the modulation amplitudes $\alpha_1/\alpha_2$ is varied, obtaining measured values of $230$ kHz and $250$ kHz at laser frequencies of $351.32020$ THz and $351.58428$ THz, respectively. 

We also investigated the effect of a nonzero $\Delta\mu_1-\Delta\mu_2$ in our experiment. Unlike in Fig.~2(b) of the main text, the modulation frequency is varied simultaneously to keep $\delta=0$ when scanning the magnetic field. This yields an extracted half-distance of $13$ kHz, which is slightly broadened compared to the value of $11$ kHz in Fig.~2(b).

\subsection{Verification of the BIC Condition}
We now verify the BIC condition in Fig.~\ref{sup8} and Fig.~\ref{sup9} using the measured parameters. The parameters used in this calculation are listed below.

The differential modulation amplitude between the $4g4$ state and the free-atom state is 200 kHz, that between the $6s$ state and the free-atom state ranges from $-26~\mathrm{kHz}$ to $-22~\mathrm{kHz}$ over the magnetic field range of 20.25-20.40 G in Fig.~\ref{sup8} and Fig.~\ref{sup9}, and that between the $4g4$ state and the $6s$ state is thus between $222~\mathrm{kHz}$ and $226~\mathrm{kHz}$. The resulting $\Gamma_{10}^{\rm eff}$ and $\Gamma_{20}^{\rm eff}$ are estimated as $\Gamma_{10}^{\rm eff}=\frac{150a_0\times7 \mathrm{kHz}}{a_{bg}}$ and $\Gamma_{20}^{\rm eff}\sim0.005(1) \Gamma_{10}^{\rm eff}$ with $a_{bg}\Gamma_{20}\sim2\times10^8~a_0\mathrm{kHz}$ (an average value), respectively. $\left| V_{12}^{\rm eff}+G_{12}^{\rm eff} \right|$ is estimated to be 11 kHz based on the experimental data in the previous subsection. Then BIC occurs at $\Delta-\delta=\left| V_{12}^{\rm eff}+G_{12}^{\rm eff} \right|\frac{\Gamma_{10}^{\rm eff}-\Gamma_{20}^{\rm eff}}{\sqrt{\Gamma_{10}^{\rm eff}\Gamma_{20}^{\rm eff}}}\sim157(16) \mathrm{kHz}$. This indicates that one of the Fano branches disappears near a magnetic field value of $155 \mathrm{kHz}/\Delta\mu_1+B_0\sim20.29(2)~\mathrm{G}$, with $\Delta\mu_1=0.797$ MHz/G and $B_0=20.09$ G. In Fig.~\ref{sup8}, the narrow Fano branch is strongly suppressed near this value. Since $\Gamma_{20}^{\rm eff}$ is much smaller than $\Gamma_{10}^{\rm eff}$, the Fano profiles as a function of modulation frequency at a fixed magnetic field in Fig.~\ref{sup9} are also suppressed near this magnetic field value.

\subsection{Fano Center and Transparency Peak}
\begin{figure*}[h]
    \centering
    \includegraphics[width=1\textwidth]{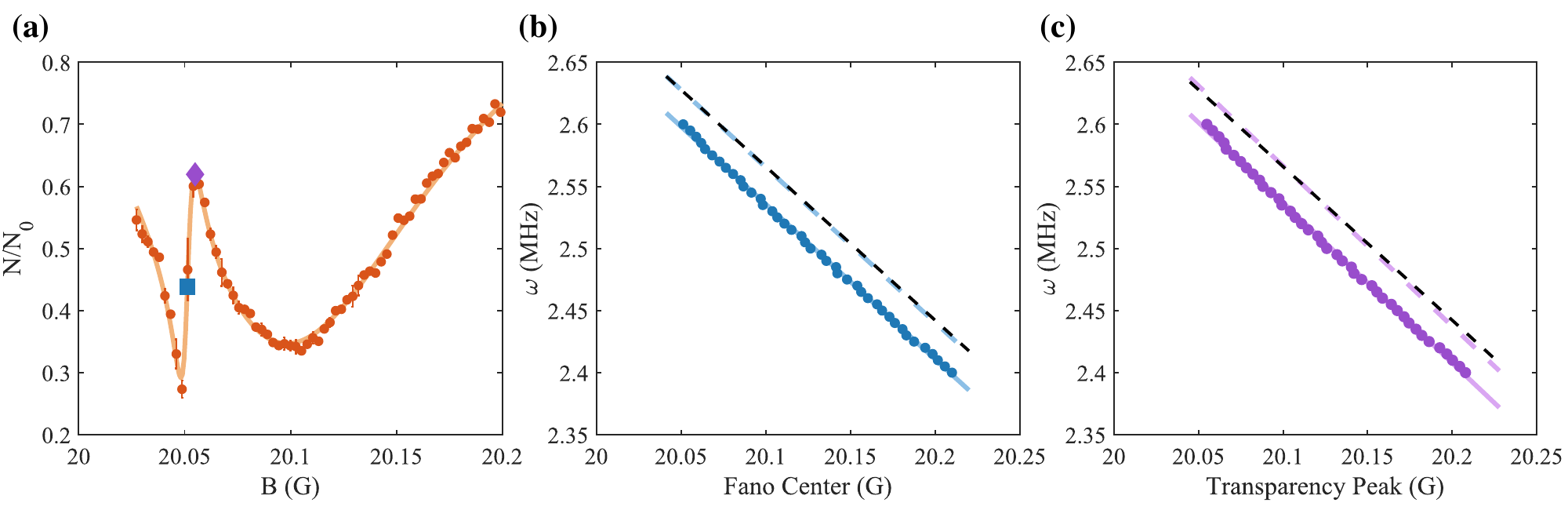}
    \caption{\textcolor{black}{ Comparision of Fano center and transparency peak. (a) Magnetic-field loss spectrum at the modulation frequency of  \(\omega=2\pi\times 2.600\)~MHz. The profile is fitted with a sum of two Fano line shapes. The blue square denotes the Fano center of the narrow branch, while the purple diamond denotes the peak position of the fitted profile. (b) Fano Center positions extracted from Fano-type fits to the magnetic-field spectra at each modulation frequency using the data in Fig.~3(a) of the main text. (c) Transparency peak positions fits to the magnetic-field spectra using the data in Fig.~3(a) of the main text. The blue (purple) solid lines represent linear fits to the data, while the blue (purple) dashed lines are obtained after compensating for the light shift. The black lines denote the energy of $|6s\rangle$ measured by modulation spectroscopy in Ref.~\cite{Ref1}.}} 
    \label{6s}
\end{figure*}

As discussed in the main text, the transparency curve in Fig.~3(a) is characterized at each modulation frequency by the fitted magnetic field center of the narrow Fano branch. An alternative quantification is given by the transparency peak, defined as the magnetic field that maximizes the survival fraction at each modulation frequency. Figure~\ref{6s}(a) illustrates these two definitions: the blue square is the fitted center of the narrow Fano branch, and the purple diamond is the extracted peak position of the fitted Fano profiles. Generally, the Fano center and the peak position do not completely coincide but remain close to each other. As shown in (b) and (c), both extracted positions are well described by a linear fit. After compensating for the light shift, the corrected Fano center line (blue dashed line in (b)) agrees well with the measured $6s$ binding energy line (black dot-dashed line in (b) and (c)), whereas the transparency peak line (purple dashed line in (c)) shows a slight deviation. This behavior is consistent with the discussion in Section~\ref{section3}, where the loss minima are slightly shifted by weak residual \(|a\rangle\leftrightarrow |m_2\rangle\) coupling and nonzero $\gamma_2$, whereas the Fano center remains less affected.

\subsection{Scaling of Transparency Width}
As discussed in the main text, the transparency width \(\Delta_{\mathrm{MWIT}}\) scales linearly with $\frac{I^2}{\Delta_{b}}$, where $\Delta_{b}$ is the Fano width of the measured background loss associated with the first molecular state. This is consistent with the expression $\left(2g \right)^2/\gamma_1+\gamma_2$ derived in Section~\ref{section3}, provided that the Fano width $\Delta_{b}$ fitted from the magnetic-field spectra is proportional to $\gamma_1$ and that $\gamma_2$ is constant. To verify that $\gamma_2$ varies slowly with increasing modulation intensity, we further fit the data using the function $a\frac{I^2}{\Delta_{b}}+c_1I+c_2$ and find that $c_1$ is negligible in our experiment.

\section{MEAN-FIELD DESCRIPTION AND  COHERENT ATOMIC AND MOLECULAR CONDENSATE DYNAMICS}

Another natural description of matter wave induced transparency in BECs is to treat atoms and molecules as coherent matter fields and investigate their equations of motion under an interacting second-quantized Hamiltonian \cite{Mean1,Mean2,Mean3}. In contrast to the stationary scattering formulation in the previous section, this approach enables the study of coherent atom-molecule oscillations, atom-molecule and molecule-molecule collisions, as well as many-body collision dynamics \cite{Ref3,Mean1,Mean2,Mean3,App1,App2}. As in previous studies of three-level or three-mode atom-molecule systems \cite{Mean2, Mean4, Mean5, YAVUZ2004253}, we also adopt the mean-field approximation by replacing field operators with $c$-numbers. However, unlike in previous works, the coupling between all levels in our scheme originates from intrinsic collision interactions and is effectively controlled via modulation parameters. We now illustrate this with the simplest case.


We denote the atomic field by \(\psi_{a}\), and the two molecular fields in the closed channel by \(\psi_{m_1}\) and \(\psi_{m_2}\). Within the mean-field treatment, the equations of motion for the field operators are given by

\begin{equation}
i\hbar\frac{\partial \psi_{a}}{\partial t}=V_{a}\psi_{a}+\left(\sum_k g_{a,k}|\psi_k|^2\right)\psi_{a}+2\alpha^*\hbar\psi_{a}^*\psi_{m_1}+2\beta^*\hbar\psi^*_a\psi_{m_2},
\label{eq:MF1}
\end{equation}

\begin{equation}
i\hbar\frac{\partial \psi_{m_1}}{\partial t}
+
i\hbar\frac{\gamma_1}{2}\psi_{m_1}
=
\left[V_{m_1}+\hbar A\cos(\omega t)\right]\psi_{m_1}
+
\left(\sum_k g_{m_1,k}|\psi_k|^2\right)\psi_{m_1}
+
\alpha\hbar\psi_{a}^2
+
\frac{\hbar\Omega}{2}\psi_{m_2},
\label{eq:MF2}
\end{equation}
and
\begin{equation}
i\hbar\frac{\partial \psi_{m_2}}{\partial t}
+
i\hbar\frac{\gamma_2}{2}\psi_{m_2}
=
V_{m_2}\psi_{m_2}
+
\left(\sum_k g_{m_2,k}|\psi_k|^2\right)\psi_{m_2}
+\beta\hbar\psi_a^2
+\frac{\hbar\Omega^*}{2}\psi_{m_1}.
\label{eq:MF3}
\end{equation}
Here, $k\in \{a,m_1,m_2\}$, \(\hbar A\cos(\omega t)\) is the modulation-induced energy shift of the first molecular state, \(\gamma_1\) and \(\gamma_2\) are the decay rates of the two molecular fields, and \(V_k\) are the uniform potentials seen by the corresponding fields. 
The parameters \(\alpha\) and \(\beta\) describe the coherent coupling between two atoms and one molecule in the states \(|m_1\rangle\) and \(|m_2\rangle\), respectively. 
The parameter \(\Omega\) is defined as the Rabi frequency of intrinsic coupling between two molecular states, which arises from interatomic interactions such as exchange and van der Waals interactions, as well as magnetic dipole interaction and second-order spin-orbit coupling.
The coefficients \(g_{jk}\) describe the mean-field interactions between the fields and are related to the corresponding background scattering lengths.

We now transform to a rotating frame that removes the explicit energy modulation of the first molecular state and shifts the second molecular state by one modulation quantum:
\begin{equation}
\phi_{a}=\psi_{a},\qquad
\phi_{m_1}=\psi_{m_1}e^{i(A/\omega)\sin(\omega t)},\qquad
\phi_{m_2}=\psi_{m_2}e^{-i\omega t}.
\label{eq:MF_transform}
\end{equation}
Using the Jacobi--Anger expansion
\begin{equation}
e^{i(A/\omega)\sin(\omega t)}
=
\sum_n J_n(A/\omega)e^{in\omega t},
\end{equation}
and keeping only the near-resonant terms, we obtain
\begin{equation}
i\hbar\frac{\partial \phi_{a}}{\partial t}
=
V_{a}\phi_{a}
+
\left(\sum_k g_{a,k}|\phi_k|^2\right)\phi_{a}
+
2\alpha^*J_0(A/\omega)\hbar\phi_{a}^*\phi_{m_1},
\label{eq:MF7}
\end{equation}
\begin{equation}
i\hbar\frac{\partial \phi_{m_1}}{\partial t}
+
i\hbar\frac{\gamma_1}{2}\phi_{m_1}
=
V_{m_1}\phi_{m_1}
+
\left(\sum_k g_{m_1,k}|\phi_k|^2\right)\phi_{m_1}
+
\alpha J_0(A/\omega)\hbar\phi_{a}^2
-
\frac{\hbar\Omega_1}{2}\phi_{m_2},
\label{eq:MF8}
\end{equation}
and
\begin{equation}
i\hbar\frac{\partial \phi_{m_2}}{\partial t}
+
i\hbar\frac{\gamma_2}{2}\phi_{m_2}
=
\left(V_{m_2}+\hbar\omega\right)\phi_{m_2}
+
\left(\sum_k g_{m_2,k}|\phi_k|^2\right)\phi_{m_2}
-
\frac{\hbar\Omega_1^*}{2}\phi_{m_1},
\label{eq:MF9}
\end{equation}
where
\begin{equation}
\Omega_1=\Omega J_1(A/\omega)
\end{equation}
is the modulation-induced effective coupling between the two molecular fields. 
The minus sign comes from \(J_{-1}(x)=-J_1(x)\).

Using Eqs.~\ref{eq:MF7}–\ref{eq:MF9}, one can thus simulate the coherent dynamical evolution of atoms and molecules through numerical solutions under specific parameters. In addition, density-dependent decay of atomic and molecular condensates arising from atom–molecule collisions and molecule–molecule
collisions can be incorporated, for example, by adding $i\hbar\gamma_{a,m_1}\left|\phi_{m_1} \right|^2\phi_a$ to the left-hand side of Eqs.~\ref{eq:MF7} and $i\hbar\gamma_{a,m_1}\left|\phi_{a} \right|^2\phi_{m_1}$ to that of Eq.~\ref{eq:MF8}.

To obtain an analytic expression of the scattering length, we consider a uniform gas and transform it to an interaction picture. 
We define
\begin{equation}
\phi_{a}(t)=\varphi_{a}(t)e^{-i\omega_0 t},\qquad
\phi_{m_1}(t)=\varphi_{m_1}(t)e^{-i\omega_1 t},\qquad
\phi_{m_2}(t)=\varphi_{m_2}(t)e^{-i\omega_2 t},
\end{equation}
where the frequencies include the static single-particle energies and the initial mean-field shifts,
\begin{equation}
\hbar\omega_0
=
V_{a}
+
\sum_k g_{a,k}|\varphi_k(0)|^2,
\end{equation}
\begin{equation}
\hbar\omega_1
=
V_{m_1}
+
\sum_k g_{m_1,k}|\varphi_k(0)|^2,
\end{equation}
and
\begin{equation}
\hbar\omega_2
=
V_{m_2}
+
\hbar\omega
+
\sum_k g_{m_2,k}|\varphi_k(0)|^2.
\end{equation}
The two detunings are defined as
\begin{equation}
\Delta=\omega_1-2\omega_0,
\qquad
\delta=\omega_2-2\omega_0.
\label{eq:MF_detunings}
\end{equation}
Here, \(\Delta\) is the detuning between the atomic pair and the first molecular state, while \(\delta\) is the detuning between the atomic pair and the second molecular state shifted by a single Floquet quantum.

After this transformation, the equations are
\begin{equation}
i\frac{\partial \varphi_{a}}{\partial t}
=
2\alpha^*J_0(A/\omega)\varphi_{a}^*\varphi_{m_1}
+
\sum_k\frac{g_{a,k}}{\hbar}
\left[
|\varphi_k(t)|^2-|\varphi_k(0)|^2
\right]\varphi_{a},
\label{eq:MF10}
\end{equation}
\begin{equation}
i\frac{\partial \varphi_{m_1}}{\partial t}
+
\left(-\Delta+i\frac{\gamma_1}{2}\right)\varphi_{m_1}
=
\alpha J_0(A/\omega)\varphi_{a}^2
-
\frac{\Omega_1}{2}\varphi_{m_2}
+
\sum_k\frac{g_{m_1,k}}{\hbar}
\left[
|\varphi_k(t)|^2-|\varphi_k(0)|^2
\right]\varphi_{m_1},
\label{eq:MF11}
\end{equation}
and
\begin{equation}
i\frac{\partial \varphi_{m_2}}{\partial t}
+
\left(-\delta+i\frac{\gamma_2}{2}\right)\varphi_{m_2} =
-
\frac{\Omega_1^*}{2}\varphi_{m_1}
+
\sum_k\frac{g_{m_2,k}}{\hbar}
\left[
|\varphi_k(t)|^2-|\varphi_k(0)|^2
\right]\varphi_{m_2}.
\label{eq:MF12}
\end{equation}

We then focus on the slow atomic dynamics and adiabatically eliminate the molecular fields. After neglecting the residual mean-field frequency shifts and setting \(\partial_t\varphi_{m_1}\simeq 0\) and \(\partial_t\varphi_{m_2}\simeq 0\), we have
\begin{equation}
\varphi_{m_1}
=
\frac{
2\alpha J_0(A/\omega)(-2\delta+i\gamma_2)
}
{
(-2\Delta+i\gamma_1)(-2\delta+i\gamma_2)-|\Omega_1|^2
}
\varphi_{a}^2,
\label{eq:MF_phi1}
\end{equation}
and
\begin{equation}
\varphi_{m_2}
=
-
\frac{
2\alpha \Omega_1^* J_0(A/\omega)
}
{
(-2\Delta+i\gamma_1)(-2\delta+i\gamma_2)-|\Omega_1|^2
}
\varphi_{a}^2.
\label{eq:MF_phi2}
\end{equation}
Substituting Eq.~\eqref{eq:MF_phi1} into Eq.~\eqref{eq:MF10}, the atomic field obeys an effective nonlinear equation
\begin{equation}
i\frac{\partial \varphi_{a}}{\partial t}
=
\frac{
4|\alpha J_0(A/\omega)|^2(-2\delta+i\gamma_2)
}
{
(-2\Delta+i\gamma_1)(-2\delta+i\gamma_2)-|\Omega_1|^2
}
|\varphi_{a}|^2\varphi_{a}.
\label{eq:MF_atom}
\end{equation}
Comparing this expression with the usual mean-field equation
\begin{equation}
i\frac{\partial \varphi_{a}}{\partial t}
=
\frac{4\pi\hbar}{m}a|\varphi_{a}|^2\varphi_{a},
\end{equation}
we obtain the effective scattering length:
\begin{equation}
a
=
a_{bg}
-
\frac{m}{2\pi\hbar}
\frac{
|\alpha J_0(A/\omega)|^2
}
{
(\Delta -i\gamma_1/2)
-
|\Omega_1/2|^2/(\delta-i\gamma_2/2)
},
\label{eq:MF_a}
\end{equation}
which recovers the expression in Section~\ref{three-channel}.